\documentclass{aa}
\usepackage{graphicx,amssymb}

\begin{document}


\title{The FORS Deep Field Spectroscopic Survey
\thanks{Based on observations obtained with FORS at the VLT, Paranal, Chile
on the course of the observing proposals 63.O-0005, 64.O-0149, 64.O-0158, 
65.O-0049, 66.A-0547, 68.A-0013, 68.A-0014, 69.A-0104}}

\author{S.\,Noll\inst{1}
\and D.\,Mehlert\inst{1}
\and I.\,Appenzeller\inst{1}
\and R.\,Bender\inst{2}
\and A.\,B\"ohm\inst{3}
\and A.\,Gabasch\inst{2}
\and J.\,Heidt\inst{1}
\and U.\,Hopp\inst{2}
\and K.\,J\"ager\inst{3}
\and S.\,Seitz\inst{2}
\and O.\,Stahl\inst{1}
\and C.\,Tapken\inst{1}
\and B.L.\,Ziegler\inst{3}}

\offprints{S.\,Noll,\\
\email{snoll@lsw.uni-heidelberg.de}}

\institute{Landessternwarte Heidelberg, K\"onigstuhl,
  D-69117 Heidelberg, Germany
\and Universit\"atssternwarte M\"unchen, Scheinerstra{\ss}e~1,
  D-81679 M\"unchen, Germany
\and Universit\"ats-Sternwarte G\"ottingen, Geismarlandstra{\ss}e~11,
  D-37083 G\"ottingen, Germany}

\date{Received; accepted}

\abstract{We present a catalogue and atlas of low-resolution spectra of a 
well defined sample of 341 objects in the FORS Deep Field. All spectra were 
obtained with the FORS instruments at the ESO VLT with essentially the same 
spectroscopic set-up. The observed extragalactic objects cover the redshift 
range $0.1$ to $5.0$. 98 objects are starburst galaxies and QSOs at $z > 2$. 
Using this data set we investigated the evolution of the characteristic 
spectral properties of bright starburst galaxies and their mutual relations 
as a function of redshift. Significant evolutionary effects were found for 
redshifts $2 < z < 4$. Most conspicuous are the increase of the average 
C\,IV absorption strength, of the dust reddening, and of the intrinsic UV 
luminosity, and the decrease of the average Ly$\alpha$ emission strength with 
decreasing redshift. In part the observed evolutionary effects can be 
attributed to an increase of the metallicity of the galaxies with cosmic age. 
Moreover, the increase of the total star-formation rates and the stronger 
obscuration of the starburst cores by dusty gas clouds suggest the occurrence 
of more massive starbursts at later cosmic epochs.

}

\authorrunning{Noll et al.} 

\maketitle

\section{Introduction}\label{introduction}

The advent of the 10\,m-class telescopes has allowed direct access to the 
early stages of galaxy evolution using spectroscopic methods. Observing with 
Keck, Steidel et al. (\cite{STE96a}, \cite{STE96b}) and Lowenthal et al. 
(\cite{LOW97}) succeeded first in acquiring spectra of high-redshift galaxies 
at $z \sim 3$. Part of the first objects were selected from the Hubble Deep 
Field North (Williams et al. \cite{WIL96}; review by Ferguson et al. 
\cite{FERG00}). All candidates were identified using a two-colour selection 
method based on the Lyman-limit break. In the following years many more 
`Lyman-break galaxies' were observed at redshifts $z \sim 3$ (e.g. Cristiani 
et al. \cite{CRI00}; Vanzella et al. \cite{VAN02}; Steidel et al. 
\cite{STE03}), $z \sim 4$ (Steidel et al. \cite{STE99}) and higher (Lehnert 
\& Bremer \cite{LEH03}). An extended review of this topic has been presented 
by Giavalisco (\cite{GIA02}).

In addition to the two-colour diagram method, which is biased towards objects 
with high star-formation rates and/or strong intergalactic absorption, deep 
multi-band photometric surveys resulting in `photometric redshifts' (Connolly 
et al. \cite{CON97}; Fern\'andez-Soto et al. \cite{FERN99}) have also been 
used to identify high-redshift galaxies. In this way, e.g., candidates with 
$1.5 < z < 2.5$ (Savaglio et al. \cite{SAV04}; Daddi et al. \cite{DAD04}) and 
$z > 5$ (Spinrad et al. \cite{SPI98}; Weymann et al. \cite{WEY98}) could be 
successfully identified. Finally, objects with strong Ly$\alpha$ emission and 
a weak continuum can be identified using narrow-band filters (Hu \& McMahon 
\cite{HU96}; Cowie \& Hu \cite{COW98}; Hu et al. \cite{HU99}; Rhoads et al. 
\cite{RHO03}; Maier et al. \cite{MAI03}). 
     
Because of the faintness of high-redshift galaxies, detailed spectral studies
have so far been restricted mainly to composite spectra (Lowenthal et al. 
\cite{LOW97}; Steidel et al. \cite{STE01}; Shapley et al. 2003, \cite{SHA03} 
hereafter) or to galaxies amplified by gravitational lensing (Pettini et al. 
\cite{PET00}, \cite{PET02}; Mehlert et al. \cite{MEH01}; Frye et al. 
\cite{FRY02}; Hu et al. \cite{HU02}).

The various spectroscopic studies of Lyman-break galaxies have shown that
these objects are vigorously star-forming galaxies having large-scale 
outflows of neutral gas. Typical outflow velocities of a few 100\,km/s were
derived from the blue-shift of the interstellar absorption lines (e.g. 
Pettini et al. \cite{PET00}; Adelberger et al. \cite{ADE03}). Investigations
of the shape of the Ly$\alpha$ line and its relation to other spectral 
properties (see \cite{SHA03}) revealed a complex dependence of the 
Ly$\alpha$ emission strength on the (neutral) gas and dust distribution and 
kinematics in these galaxies. Pettini et al. (\cite{PET00}, \cite{PET02}) 
using weak low-ionisation lines and Mehlert et al. (\cite{MEH02}) using the 
high-ionisation stellar-wind blend C\,IV\,$\lambda\lambda\,1548,1550$ find
evidence for a chemical evolution with cosmic age of the bright starburst 
galaxies. Further indications for evolutionary effects come from photometric 
data on the luminosity, size and mass of high-redshift galaxies (e.g. 
Lowenthal et al. \cite{LOW97}; Madau et al. \cite{MAD98}; Steidel et al. 
\cite{STE99}; Shapley et al. \cite{SHA01}; Ferguson et al. \cite{FERG04}; 
Idzi et al. \cite{IDZ04}), suggesting the formation of more massive galaxies 
at later cosmic epochs.     
  
In the present paper we describe a deep spectroscopic survey using 
high-quality low-resolution spectra of galaxies covering the redshift range 
$0 < z < 5$. This survey allows a detailed analysis of the dependence of 
basic galaxy properties on cosmic age. We primarily focus on the analysis of 
evolutionary effects in the redshift range $2 < z < 4$, where we expected 
the most significant results. A detailed discussion of other redshift ranges 
will be presented elsewhere. 

This survey is part of the FORS Deep Field (FDF) project (Appenzeller et al. 
\cite{APP00}; Heidt et al. \cite{HEI03b}) which has been carried out at the 
ESO Very Large Telescope (VLT) as a guaranteed time programme using the two 
FORS instruments. The FDF is located close to the South Galactic Pole and has 
a size of about $7' \times 7'$. The photometry is based on deep images in 
nine filter bands from $U$ to $K$ (Heidt et al. \cite{HEI03b}; Gabasch et al. 
\cite{GAB04}), which allows an efficient selection of candidates for 
spectroscopy using photometric redshifts. The 50\% completeness limits in the 
Vega system are $27.7$, $26.9$, $26.7$, $26.4$\,mag in $B$, $g$, $R$, and 
$I$, respectively. 

In the following we describe the FDF spectroscopic survey, discussing the 
sample selection (Sect.~\ref{sample}), the observations 
(Sect.~\ref{observations}), the data reduction (Sect.~\ref{reduction}) 
and the derivation of redshifts and object types (Sect.~\ref{analysis}). 
The catalogue of the FDF spectroscopic sample (available in electronic form 
only) is described in Sect.~\ref{catalogue}. Basic characteristics of the 
spectra and the spectroscopic redshift distribution are outlined in 
Sect.~\ref{propspec} and \ref{distribution}, respectively. In 
Sect.~\ref{specsample} we analyse the FDF high-redshift sample with regard 
to evolutionary effects. Implications from the observational results are 
discussed in Sect.~\ref{implications}. 

Throughout the paper $H_0 = 67\,{\rm km}\,{\rm s}^{-1}\,{\rm Mpc}^{-1}$, 
$\Omega_{\Lambda} = 0.7$ and $\Omega_{\rm M} = 0.3$ are adopted.

\section{The spectroscopic sample}\label{sample}

\begin{figure}
\centering 
\includegraphics[width=6cm,angle=-90]{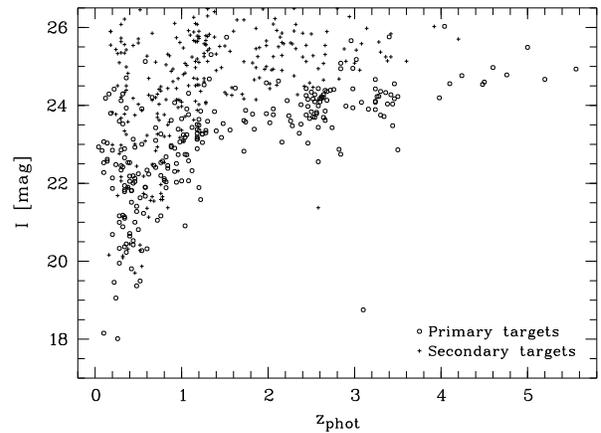}
\caption[]{$I$ magnitudes as a function of the photometric redshifts of the 
primary (circles) and secondary extragalactic targets (crosses) of the FDF 
spectroscopic survey. The bright object with $z_{\rm phot} = 3.1$ is the 
quasar Q\,0130--260 ($z_{\rm spec} = 3.365$).}
\label{fig_Iz}
\end{figure}

For the present investigation we selected a sample of galaxies in the FDF 
covering a large range of photometric redshifts derived from the FDF 
photometric programme (Heidt et al. \cite{HEI03b}). The photometric 
redshifts were calculated by fitting semi-empirical spectral energy 
distributions (SEDs) to the filter-band fluxes (Bender et al. \cite{BEN01},
\cite{BEN04}; Gabasch et al. \cite{GAB04}). The templates were constructed 
using stellar population models of Maraston (\cite{MARA98}) with variable 
reddening according to Calzetti et al. (\cite{CAL94}) and broad-band SEDs 
produced from photometric data of galaxies in the Hubble Deep Field North 
(HDF-N, Williams et al. \cite{WIL96}) with known spectroscopic redshifts. 
From a comparison of photometric and spectroscopic redshifts carried out in 
the HDF-N and the FDF we estimate for the present FDF spectroscopic sample 
a fraction of misidentifications of $\sim 3$\,\% only. The rms error of 
$z_{\rm phot} - z_{\rm spec}$ for the FDF objects with 
spectroscopically-confirmed photometric redshifts amounts to $0.13$.

To obtain sufficiently large subsamples of candidates for the different 
redshift intervals, apparent brightness limits for the object selection were 
chosen depending on the redshift. Fig.~\ref{fig_Iz} shows the $I$ magnitudes 
(Vega system) of the spectroscopically-observed objects as a function of 
photometric redshift. For the redshift range $2.0$ to $4.0$ the selection 
limit was normally $24.5$\,mag in $I$. For $I < 24$\,mag the fraction of the 
spectroscopically-observed photometric candidates was about 50\,\%. For lower 
redshifts the same level of completeness was achieved for brighter objects 
($I < 22.5$\,mag for $1.0 < z_{\rm phot} < 1.5$). For $z_{\rm phot} > 4.0$ 
one third of the objects sufficiently bright for spectroscopy ($I < 25$\,mag) 
was included. 

The majority of the targets selected are intrinsically bright objects 
($L > L^*$) at redshifts between 1 and 5. In addition to these primary 
targets the spectroscopic sample contains objects which were observed 
serendipitously as their images coincided by chance with the slit positions 
of the primary targets. Two or three additional objects on a slitlet were not 
unusual. These secondary objects constitute a random sample constrained only 
by the apparent brightness. Hence, faint intermediate-redshift galaxies 
dominate this sample, as indicated in Fig.~\ref{fig_Iz}. 

A few high-redshift galaxy candidates were selected which deviated from the 
selection criteria for $z > 2$ primary targets. Firstly, three Ly$\alpha$ 
emitters, which were initially observed as secondary targets, were converted 
to primary targets. Secondly, we selected three faint high-redshift 
candidates located in the vicinity of $z_{\rm phot} > 2$ primary objects with
$I < 24.5$\,mag. Thirdly, we included three faint objects located in the 
vicinity of the bright FDF quasar Q\,0130--260 (see Fig.~\ref{fig_qsoclust}). 
These three candidates had been selected as likely Ly$\alpha$ emitters from 
narrow-band images sensitive to Ly$\alpha$ at the quasar redshift (see Heidt 
et al. \cite{HEI03a}). Furthermore, three objects were selected on the basis
of their photometric redshifts $z_{\rm phot} \sim 3$ {\em and} a colour 
excess $g - R \ge 0.5$\,mag, indicating strong Ly$\alpha$ emission. In 
principle, these additionally selected objects increase the fraction of 
Ly$\alpha$ bright galaxies in the sample. However, the sample based on the 
standard brightness criterion is not significantly affected, since the 
continuum flux of the Ly$\alpha$ emission galaxies is generally low. Only two 
of the additional galaxies are brighter than $I = 24.5$\,mag. Consequently, 
the high-redshift sample can be considered as a representative sample for the 
absolute brightest objects with $z \ge 2$.

\section{Observations}\label{observations}

All spectroscopic observations of the FORS Deep Field were carried out using
the two FORS instruments at the ESO VLT at Paranal, Chile, mainly between
September 2000 and September 2002. The total integration time was $63.7$\,h. 

All spectra were taken using the multi-object spectroscopy modes MOS and MXU
of FORS (see FORS Manual). The MOS masks consist of 19 individually movable
and adjustable slitlets. The MXU masks are manufactured using a 
laser-cutting device. Since the MXU was not available for the first 
campaigns and only at FORS\,2, almost 90\,\% of the exposures were obtained 
using the MOS mode.

To assure a homogeneous sample the observations were carried out using a 
standardised set-up. The slit width was uniformly set to $1''$, corresponding 
approximately to the size of typical high-redshift galaxies under average 
atmospheric conditions at Paranal. For all observations discussed here, the 
low-resolution grism 150\,I was used. This grism covers the full range of the 
CCD sensitivity (3300~-- 10000\,\AA) with a relatively high efficiency, 
reaching its maximum at around 5000\,\AA{}. The grism was employed without 
order separation filters in order to exploit the whole wavelength range and 
to make most efficient use of the observing time. Since most galaxies were 
red with no or little blue flux, second order contamination in the red was 
normally negligible. In the case of blue objects with significant second 
order contamination only the uncontaminated part of the spectrum was used. 
The set-up resulted in a measured spectral resolution element of 
$\approx 23$\,\AA{} (FWHM) for the calibration lines and $\approx 18$\,\AA{} 
for point sources under good seeing conditions. 

Our high-redshift galaxies typically had an apparent brightness of 
$I > 23$~mag. On the other hand, signal-to-noise ratios of 10 and more are 
needed for quantitative spectroscopic studies. Therefore, integration times 
up to 10\,h were scheduled to achieve the desired S/N. The observations
of the individual objects were usually distributed over several mask 
set-ups to optimise the observing time allocation and to reduce systematic 
errors caused by CCD defects and object locations near slit edges. The 
individual exposures had integration times between 30 and 48\,min.

\section{Data reduction}\label{reduction}

\begin{figure}
\centering 
\includegraphics[width=6cm,angle=-90]{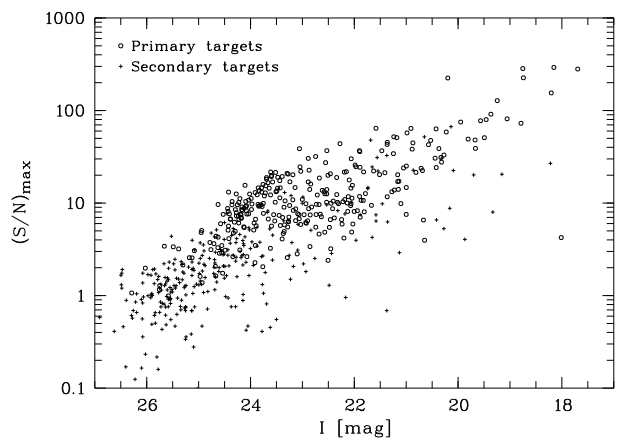}
\caption[]{Continuum signal-to-noise ratio per resolution element of the
FDF spectra (in the filter band $B$, $g$, $R$ or $I$ with the best average 
S/N) as a function of the $I$ magnitude. Circles refer to primary targets, 
crosses to secondary targets.} 
\label{fig_snI}
\end{figure}

The data were reduced using MIDAS\footnote{{\bf M}unich {\bf I}mage 
{\bf D}ata {\bf A}nalysis {\bf S}ystem} routines optimised for the
reduction of FORS data, mainly based on the MIDAS context MOS. 

For the bias subtraction for each observing run a master bias was generated 
by calculating the normalised median of a minimum of five single bias frames 
and subtracted from the raw frames after scaling the bias level with the 
overscan.

The pixel-to-pixel sensitivity variations were corrected by means of 
internal screen flat-fields provided by ESO for each mask set-up. 
Large-scale intensity fluctuations were removed by dividing the flat-fields 
by a smoothed version using a 50 pixel median. Then the multi-object spectra 
were divided by the resulting normalised flat-fields.

The wavelength calibration was carried out in two steps. First, the 
dispersion relation was derived using He, Hg/Cd and Ar spectra, which were 
obtained with the same mask configuration as the object spectra. These 
calibration spectra were used to obtain a fourth-order dispersion relation 
for each row of the MOS frame. Typical relative uncertainties of the 
relation were $\sim 0.5$\,\AA{} ($\sim 1/10$~pixel). 

The calibration spectra were obtained in zenith position during daytime. 
Therefore, bending and the positioning of the laser-cut masks in the mask 
exchange unit (MXU) could cause variations of the slitlet locations between 
calibration frames and object spectra of up to one pixel. This effect was 
corrected by measuring the position of the prominent telluric line 
[O\,I]\,$\lambda\,5577$ and by shifting the spectra accordingly. The final 
error of the wavelength calibration was found to be about $0.5$ to 1\,\AA{}.
   
The sky spectrum underlying the object spectrum was derived by interpolation 
and subtracted. The interpolation was carried out by polynomial fits of first 
to third order as well as median fits using 7 to 15 pixels. In each case the 
most suitable approach was chosen according to its success in reaching an 
undistorted object continuum compatible with the photometric data and the 
minimisation of sky line residua. The uncertainty of the zero-flux level of 
the flux calibrated one-dimensional spectra was found to correspond to a few 
$10^{-23}\,{\rm W}\,{\rm m}^{-2}\,{\rm \AA}\,\!^{-1}$ in the optical 
wavelength region.    

For the extraction of one-dimensional object spectra from the two-dimensional 
MOS spectra the S/N-optimised algorithm of Horne (\cite{HOR86}) was used. 
Pixels with a deviation greater than $5\,\sigma$ from the expected intensity 
profile were ignored in the variance-weighted summation procedure. This 
allows eliminating the majority of cosmics and CCD defects. As a by-product 
of the variance estimates the algorithm provides for each final spectrum 
$f(\lambda)$ an error function $\Delta f(\lambda)$ representing essentially 
the photon noise level. 

The atmospheric extinction correction was carried out using the ESO standard
extinction curve (T\"ug \cite{TUG77}). 

A first-order flux calibration was performed using spectra of 
spectrophotometric standard stars (Oke \cite{OKE90}; Hamuy et al. 
\cite{HAM92}, \cite{HAM94}) usually obtained each night with $5''$ wide 
slits. These spectra were reduced like the programme spectra and used to 
convert the ADUs into flux units. Then the flux-calibrated observed spectra 
were divided by the corresponding standard star spectra from the literature 
in order to obtain response curves. Eventually, representative correction 
curves were created for each run combining the individual response curves 
and smoothing the result by a spline interpolation. 

The limited slit width of $1''$ unavoidably causes flux losses in the case of 
extended objects, modest seeing and/or decentred slit positions. These flux 
losses were corrected using the total photometric fluxes measured during the 
photometric programme (Heidt el al. \cite{HEI03b}). In practice, 
quasi-photometric fluxes were calculated from the spectra (applying the known
FORS filter response curves) and divided by the photometric results to 
determine corrections. Usually the averaged factors of the $g$ and $R$ 
filters were used. For very blue or very red SEDs $g$ and $R$ were 
substituted or supplemented by $B$ or $I$.   
  
The resulting spectroscopic fluxes can still be affected by inaccuracies of
the atmospheric extinction law, the response curve, the zero-flux level and 
residua of strong sky lines. For extended objects with strong colour 
gradients the slit loss correction can be incorrect because of systematic 
deviations between the measured and the representative object spectrum. To 
estimate the error introduced by such effects, the spectroscopic and 
photometric fluxes of the final co-added spectra were compared. For the 
primary spectroscopic targets this analysis indicates typical relative flux 
uncertainties of about 5\,\% in the $g$ and $R$ range and 15\,\% in the 
$B$ and $I$ range.      

The flux-calibrated spectra were S/N-optimised co-added using the squared 
ratios
\begin{equation}\label{eq_snr}
w = \frac{\langle f(\lambda)\rangle}{\langle\Delta f(\lambda)\rangle}
\end{equation} 
of the suitably averaged spectra $f(\lambda)$ and error 
functions $\Delta f(\lambda)$ as weights. At very low S/N errors of the
spectral continuum level are no longer negligible, which may lead to 
unrealistic weights. Hence, an additional $w_0 = 0.5$ was introduced in 
Eq.~(\ref{eq_snr}), assuming that the noise levels of the individual spectra 
are of the same order. To remove artifacts such as CCD defects or undetected 
cosmics a $\sigma$-clipping procedure was carried out {\em before} the 
averaging routine was started. The clipping threshold was set to 
$6 \,\Delta f(\lambda)$.

Since the detector oversampled the spectra with about 4 pixels per resolution 
element, all spectra were smoothed to match the spectral resolution using a 
Gaussian filter. The error functions were adjusted accordingly, converting 
noise per pixel into noise per resolution element.

Finally, the spectra were corrected for Galactic dust extinction using the 
law of Cardelli et al. (\cite{CAR89}) and the expected colour excess 
$E_{B - V} = 0.018$\,mag of the extragalactic objects in the FORS Deep 
Field (Schlegel et al. \cite{SCHL98}). In the visual the correction amounts 
to about 5\,\%. 

The atmospheric B (6830 -- 6950\,\AA{}) and A (7570 -- 7710\,\AA{}) 
absorption bands were removed by dividing the spectra with a band model 
obtained from the average of a large number of high S/N spectra.\\  

The resulting data set of the FDF spectroscopic survey contains 604 co-added, 
one-dimensional spectra, 51\,\% of which represent primary targets. 
Fig.~\ref{fig_snI} shows the S/N versus $I$ magnitude for these spectra. 
Primary and secondary objects are distinguished by different symbols (circles 
and crosses). As expected, the average S/N is higher for the primary objects 
($\approx 20$) than for the secondary objects ($\approx 4$). 24\,\% of the 
spectra have a continuum S/N~$\lesssim 1$. These objects, essentially all of 
which are secondary targets, were excluded from the further analysis.

\section{Derivation of redshifts and object types}\label{analysis}

Before analysing the spectra in detail, we derived for all objects redshifts
and rough object types using the following iterative procedure:
\begin{enumerate}
\item An approximate redshift was determined from a visual line 
identification. 
\item An approximate object type was determined by comparison with known 
spectral energy distributions.
\item By averaging the spectra for each object type a set of empirical 
templates was constructed.
\item Improved redshifts and types were derived by cross-correlating the 
object spectra with these template spectra.
\item Steps (3) and (4) were repeated until an optimal fit was achieved. 
\end{enumerate} 
In detail we proceeded as follows:

\subsection{Redshifts from visually identified lines}\label{lines}

For low-redshift galaxies with lines and blends with known rest wavelength, 
redshifts were determined (preferably using strong emission lines) by 
fitting Gaussians to the line centres. Spectra of high-redshift galaxies are 
characterised by resonance lines and blends with complex profiles, such as 
blue-shifted absorption lines and P Cygni profiles (see e.g. Adelberger et 
al. \cite{ADE03}). Hence, the visual redshift derivation was based on 
features such as the emission lines He\,II\,$\lambda\,1640$ and 
C\,III]\,$\lambda\lambda\,1907,1909$ and the low ionisation absorption lines 
Si\,II\,$\lambda\,1526$, Fe\,II\,$\lambda\,1608$. These lines were expected 
to represent the systemic velocities of the galaxies or outflows not larger 
than a few 100\,km/s. Thus, our redshifts should be consistent with those of 
Pettini et al. (\cite{PET00}), Frye et al. (\cite{FRY02}), and Adelberger et 
al. (\cite{ADE03}).

\subsection{Spectral definition of object types}\label{types}

Since most of the FDF galaxies are not resolved classification schemes 
based on the morphological Hubble types as applied by Kennicutt 
(\cite{KEN92}) and Kinney et al. (\cite{KIN96}) are not useful in our case.
Therefore, we created a rough system of five galaxy spectral types (I to V) 
differing in the ratio of the rest-frame UV flux to the optical flux. The 
strengths and profiles of spectral lines have almost no impact on this
classification. Obviously a one-dimensional sequence of only five types 
cannot be sensitive to the subtleties of the composition of the stellar 
population, the influence of dust and other characteristic galaxy parameters. 
However, in view of the greatly varying S/N of our spectra and the limited 
accessible wavelength range, our classification has the advantage of being
rather robust to S/N induced effects.   

Our classification obviously requires the presence of reliable continua. 
Therefore, for the strong Ly$\alpha$ emitters with weak continua a 
supplementary template VI was used. 

QSOs (or, more generally, galaxies dominated by an active galactic nucleus), 
whose spectra could not be fitted by the spectral standard types I to V were 
allotted a special class VII.

Stars (type VIII) were identified on the basis of their angular intensity 
profiles and by comparison with stellar reference spectra from the 
literature (e.g. Pickles \cite{PIC85}). Their spectra were classified 
following the system of Morgan and Keenan (Morgan \& Keenan \cite{MOR73}; 
Keenan \& Mc\,Neil \cite{KEE76}) extended to very cool objects (e.g. 
Mart\'\i{}n et al. \cite{MART99}).

\subsection{Calculation of the template spectra}\label{templates}

\begin{figure*}
\centering 
\includegraphics[width=8cm,height=7.35cm,clip=true]{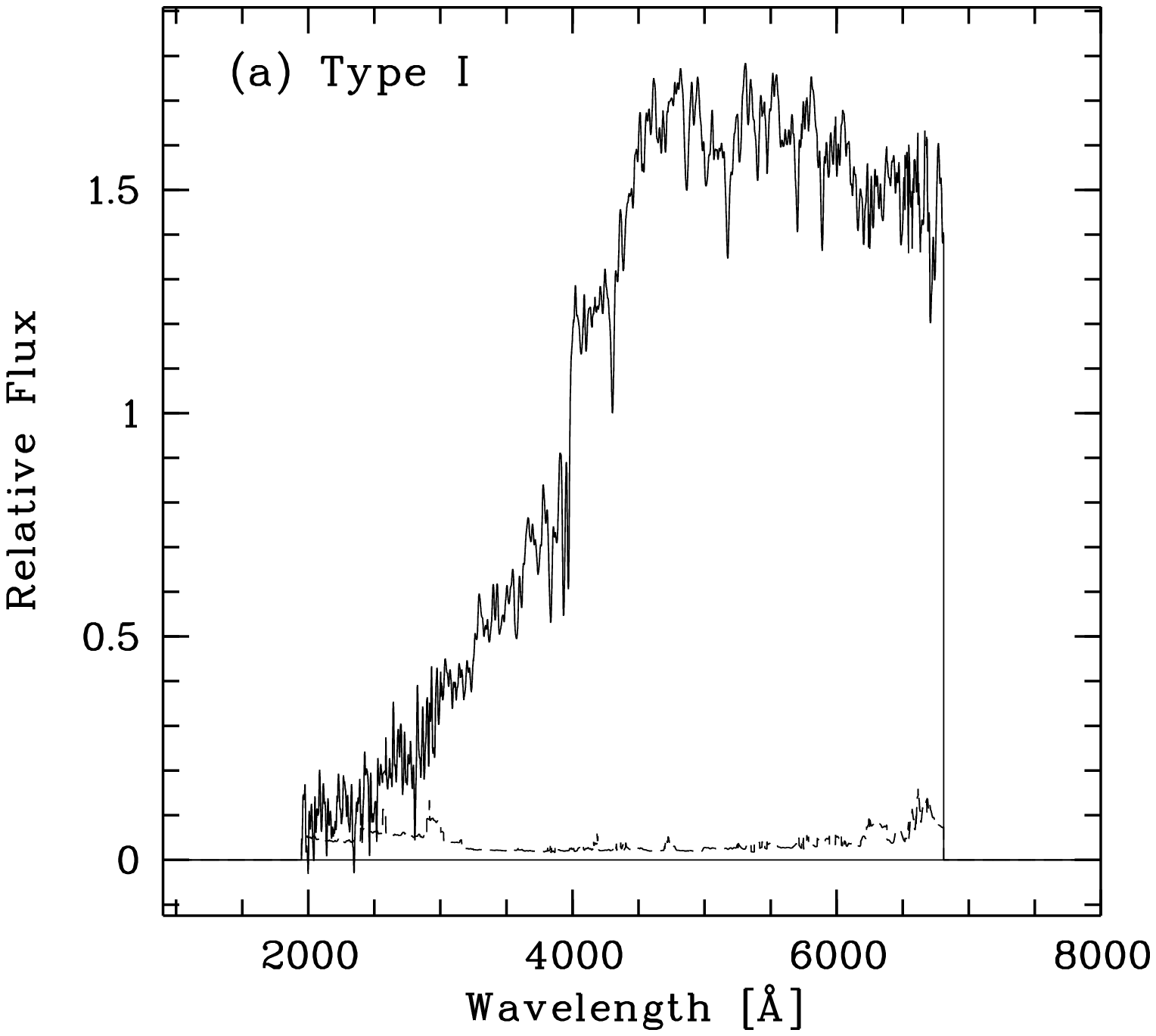}
\includegraphics[width=8cm,height=7.35cm,clip=true]{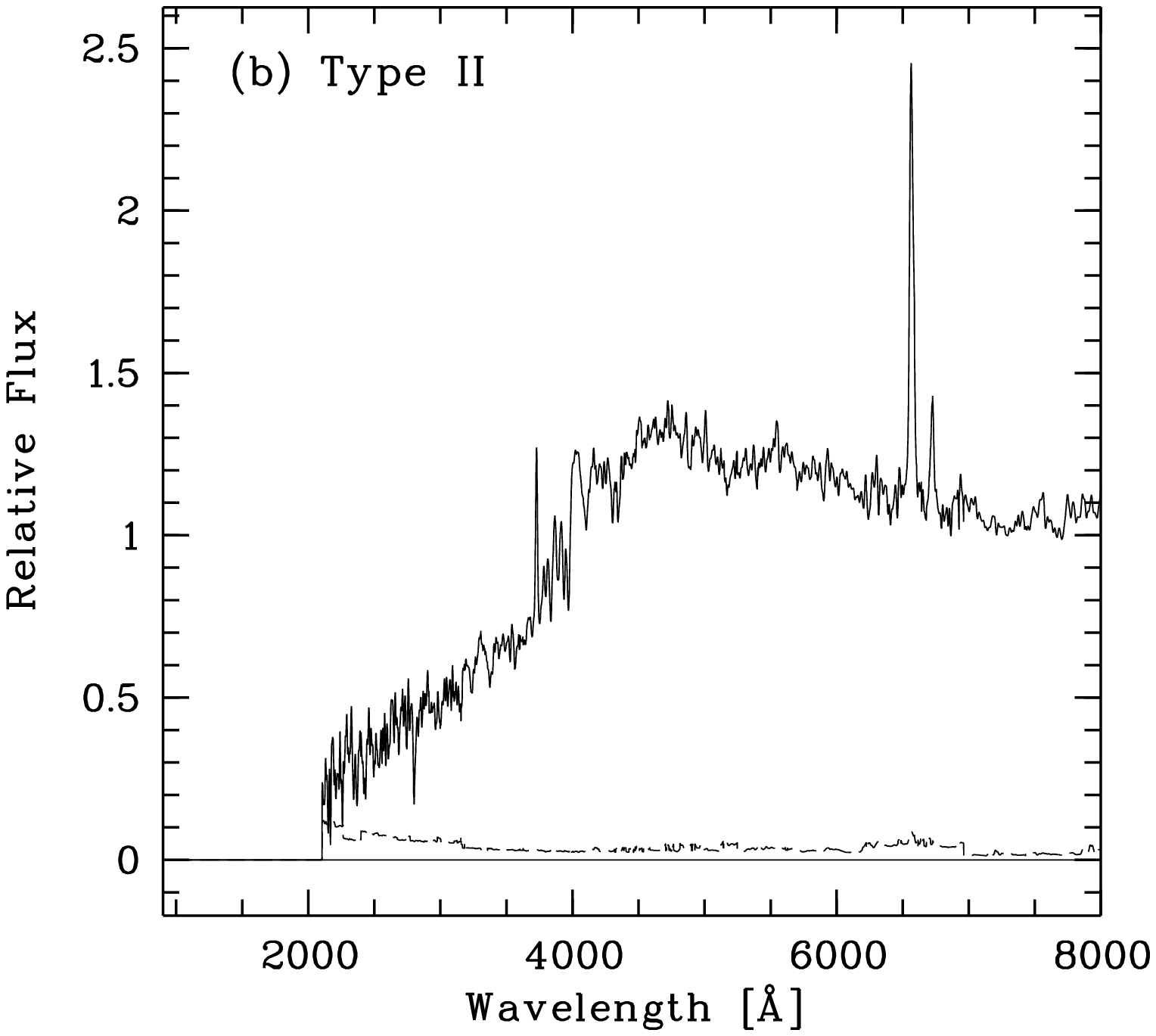} 
\includegraphics[width=8cm,height=7.35cm,clip=true]{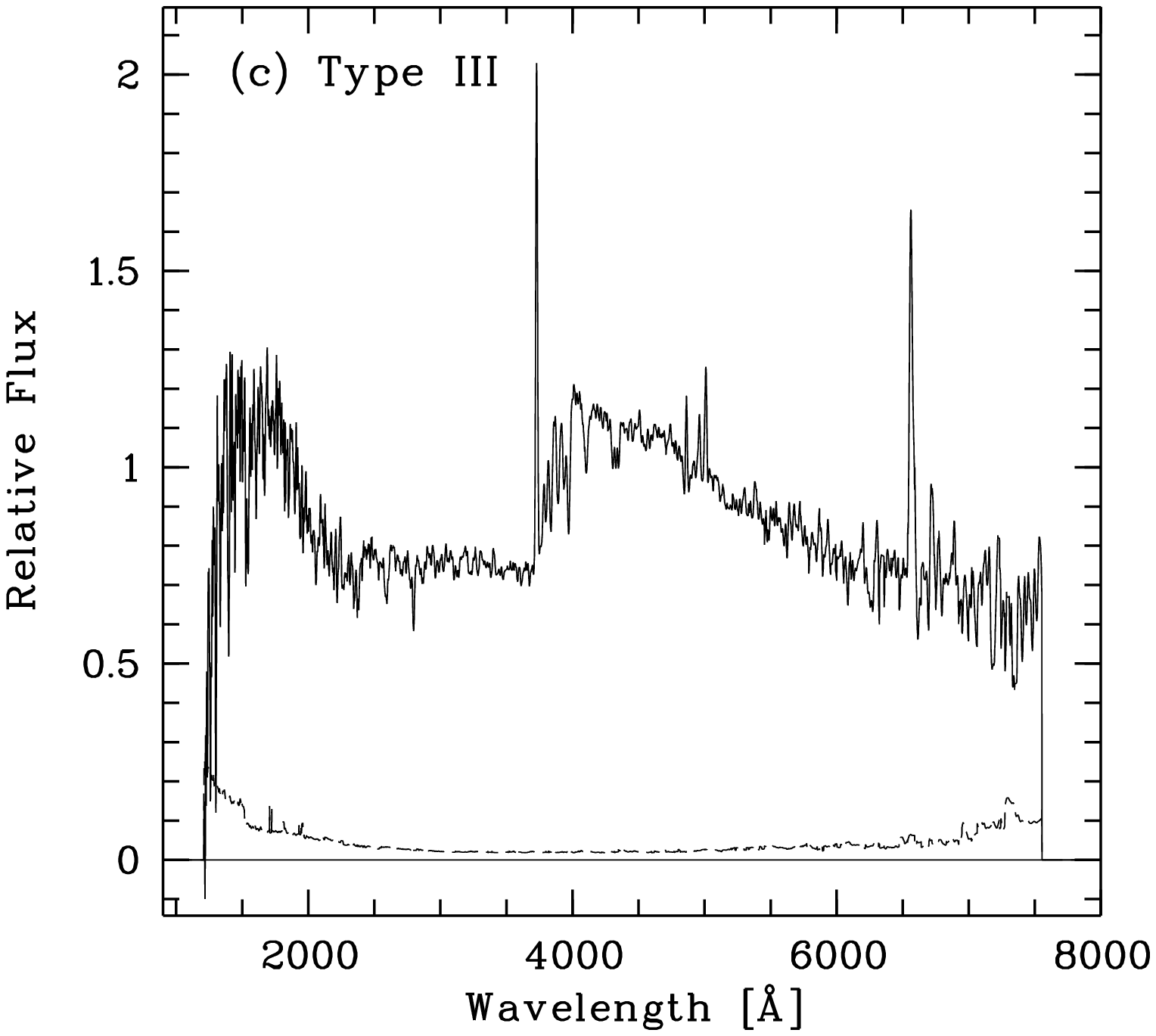} 
\includegraphics[width=8cm,height=7.35cm,clip=true]{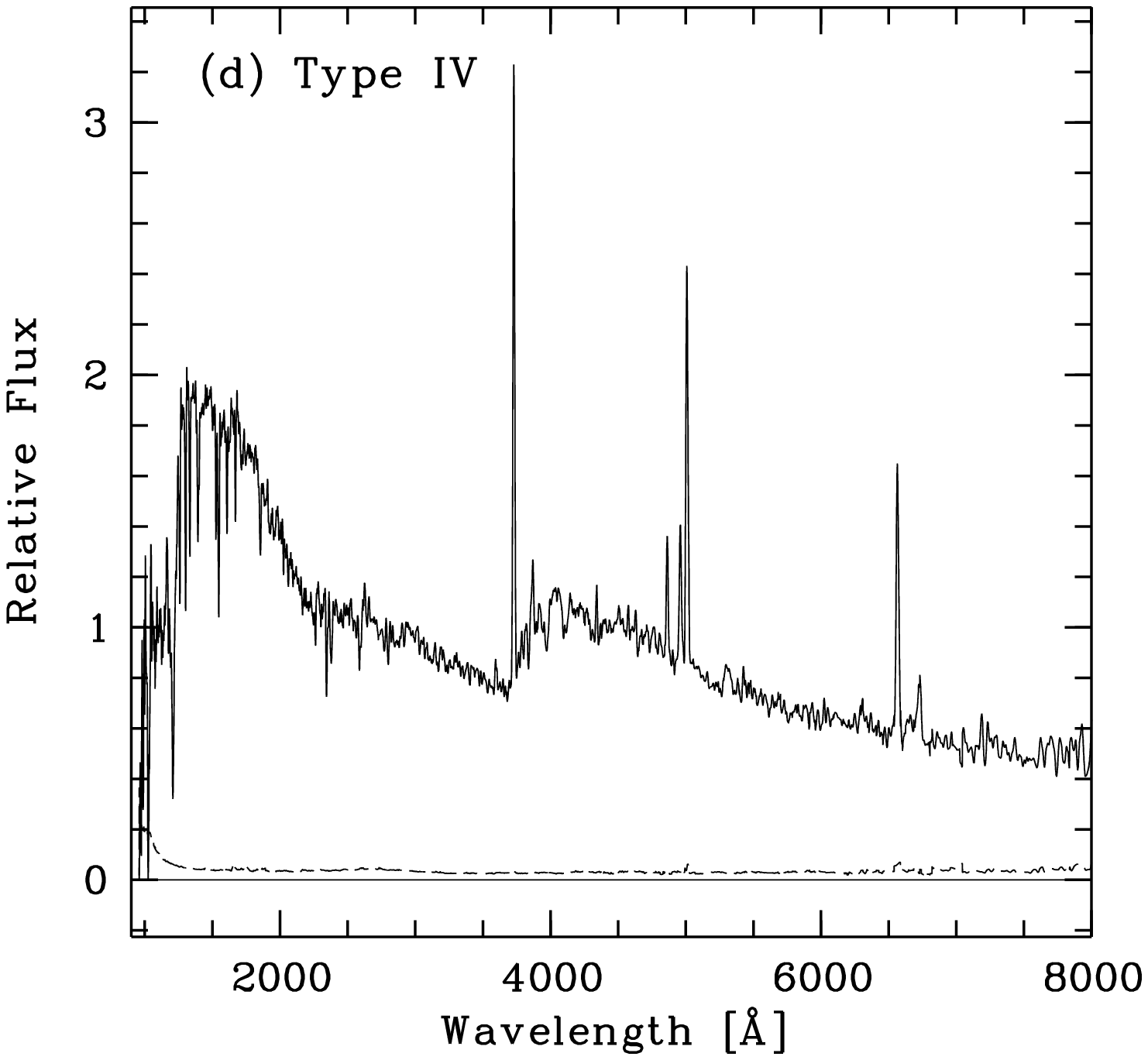}
\includegraphics[width=8cm,height=7.35cm,clip=true]{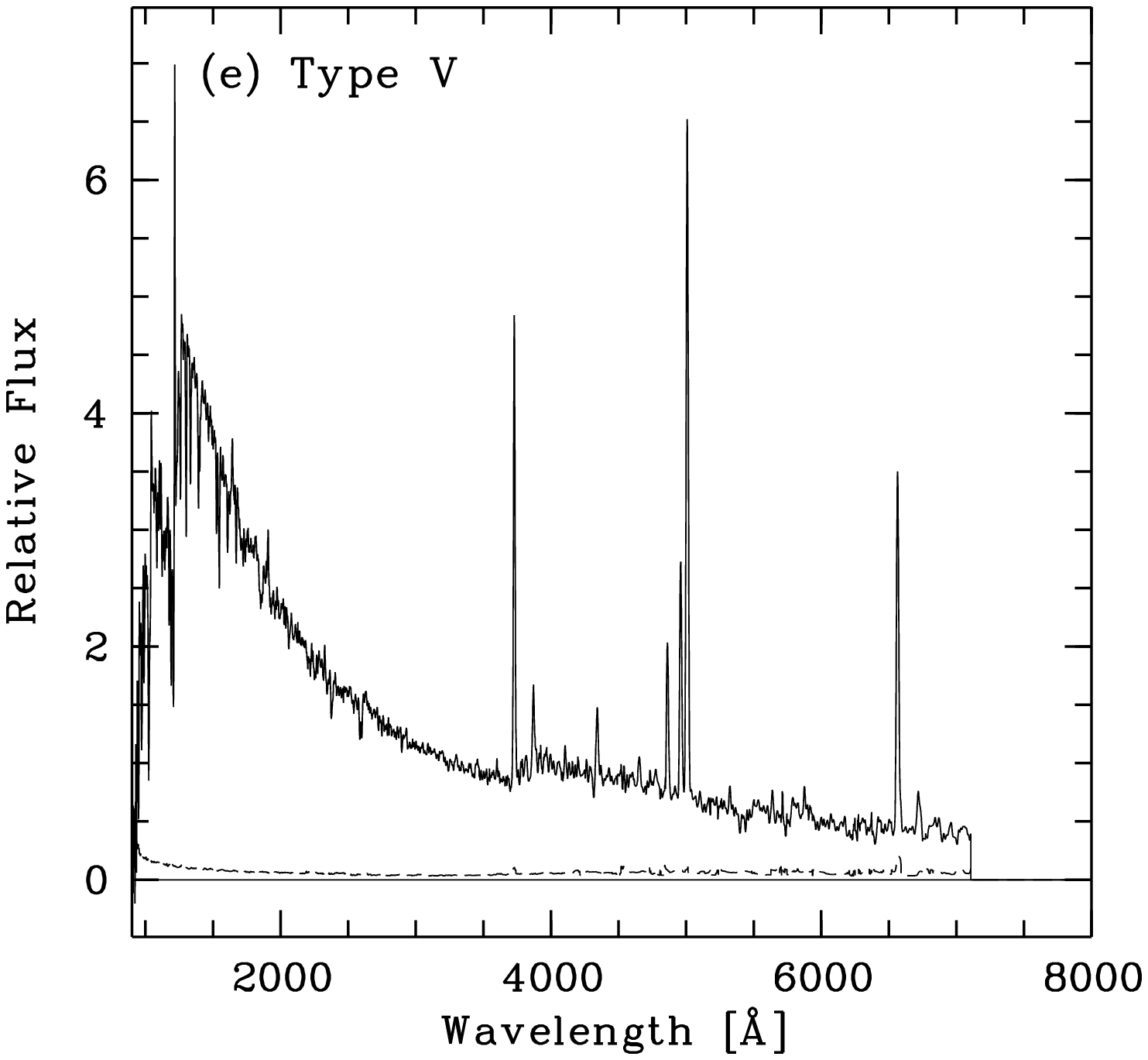}
\includegraphics[width=8cm,height=7.35cm,clip=true]{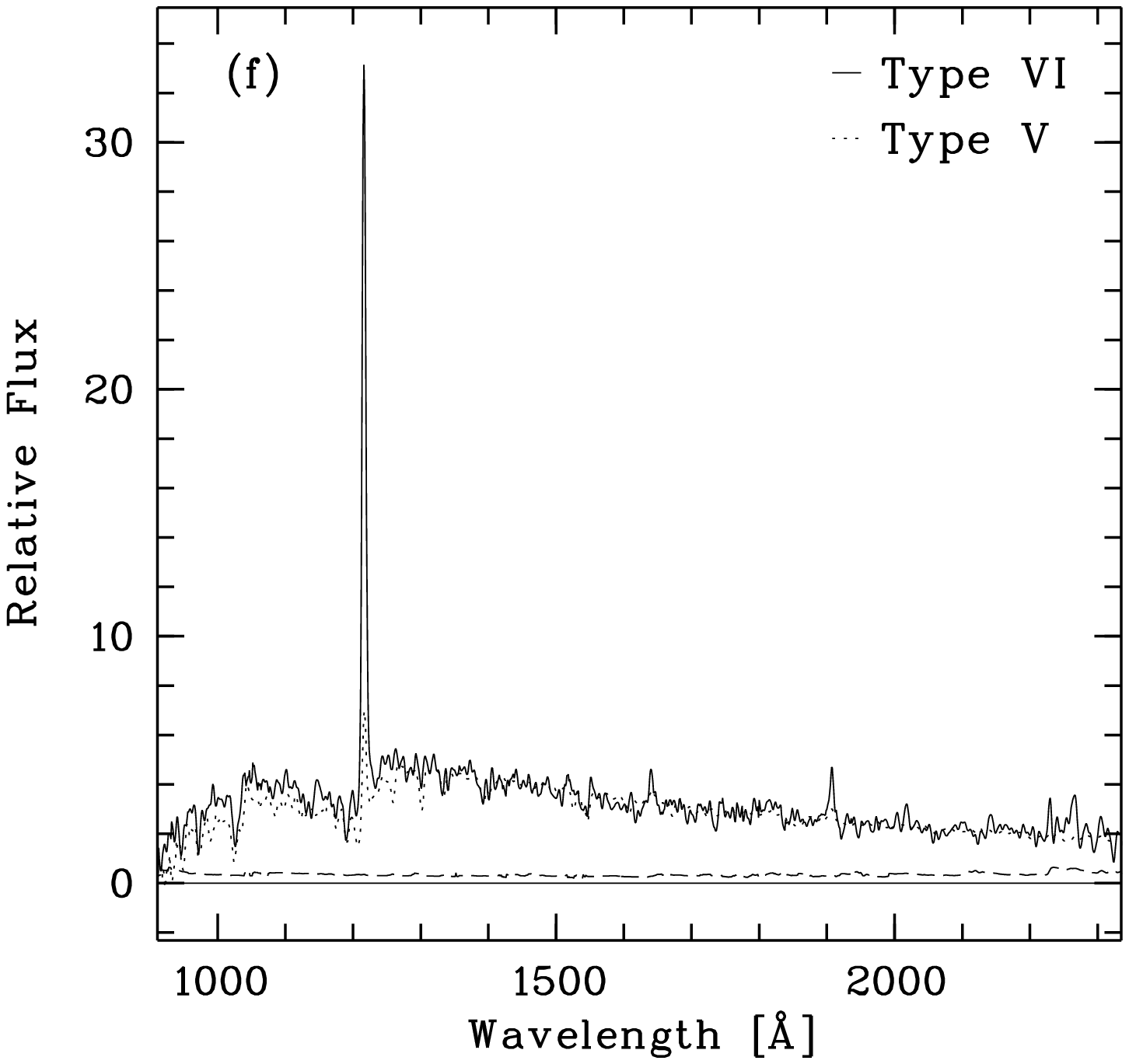}
\caption[]{Empirical templates used for classifying the galaxies of the FDF 
spectroscopic sample. The noise level is indicated by a dashed line. SEDs 
I to V are normalised to the same mean flux in the wavelength range 3500 to 
4500\,\AA{}. SED\,VI for bright Ly$\alpha$ emitters (panel~(f)) is adjusted 
to have the mean flux of SED\,V between 1400 and 1600\,\AA{}. For comparison 
SED\,V is overplotted in panel~(f) (dotted line).}
\label{fig_sed}
\end{figure*}

The template spectra were calculated by averaging FDF spectra of the 
corresponding type with secure redshifts.

As a preparatory step the individual spectra were masked to exclude 
wavelength regions affected by systematic errors. The resulting useful 
wavelength range usually included the interval between 3800 and 8250\,\AA{}, 
except for regions containing strong sky line residuals. Then the spectra 
were converted to the rest-frame system and rebinned with a step size of 
1\,\AA{}.  

To obtain average spectra with equivalent widths (EWs) equal to the average 
EW of the individual spectra, the masked rest-frame spectra 
$f_i^{\rm m}(\lambda)$ were divided into a continuum and a line part 
according to
\begin{equation}\label{eq_contline}
f_i^{\rm m}(\lambda) =  
f_{{\rm c},i}^{\rm m}(\lambda)\, f_{{\rm l},i}^{\rm m}(\lambda).
\end{equation}
The $f_{{\rm c},i}^{\rm m}(\lambda)$ were generated using a median filter of 
100\,\AA{} width.

The $f_{{\rm l},i}^{\rm m}(\lambda)$ were averaged as follows:
\begin{equation}\label{eq_avlinet}
F_{\rm l}(\lambda) = 
\frac{1}{M(\lambda)} \sum_{i = 1}^{n} f_{{\rm l},i}^{\rm m}(\lambda)
\end{equation} 
where
\begin{equation}\label{eq_summasksn}
M(\lambda) = \sum_{i = 1}^{n}\, m_i(\lambda)
\end{equation} 
is the sum of the mask functions $m_i(\lambda)$ (values either 0 or 1) of 
all $n$ spectra.

The continua were added sorted by redshift in order to maximise the 
respective size of the overlapping wavelength range. Step-by-step the 
summation was realised using
\begin{equation}\label{eq_sumcontstep}   
F_{{\rm c},i}(\lambda) = F_{{\rm c},i - 1}(\lambda) + 
\frac{f_{{\rm c},i}^{\rm m}(\lambda)}{c_i}
\end{equation}    
with $F_{{\rm c},0}(\lambda) = 0$, $c_1 = 1$ and the effective flux 
correction factor for $i \ge 2$
\begin{equation}\label{eq_fluxcorrweight}
c_i = \bigg\langle 
\frac{M_{i - 1}(\lambda)\, f_{{\rm c},i}^{\rm m}(\lambda)}
{F_{{\rm c},i - 1}(\lambda)} 
\bigg\rangle_{f_{{\rm c},i}^{\rm m}\,F_{{\rm c},i - 1} \ne 0} 
\end{equation} 
where
\begin{equation}\label{eq_summasksi}
M_i(\lambda) = \sum_{j = 1}^{i} m_j(\lambda).
\end{equation} 
$c_i$ ensures that all spectra are added with the same weight. Finally, the 
average continuum was calculated by
\begin{equation}\label{eq_avcontt}
F_{\rm c}(\lambda) = \frac{1}{M(\lambda)} F_{{\rm c},n}(\lambda)
\end{equation} 
and suitably normalised.

$F_{\rm c}(\lambda)$ was derived adding spectra in ascending as well as 
descending redshift order. The relative deviations were smaller than 5\,\%.
The final templates were calculated according to
\begin{equation}\label{eq_avupdown}
F(\lambda) = \frac{1}{2}
\big(F_{\rm c,\uparrow}(\lambda) + F_{\rm c,\downarrow}(\lambda)\big)\, 
F_{\rm l}(\lambda).
\end{equation}    

Fig.~\ref{fig_sed} shows the resulting templates for the types defined in 
Sect.~\ref{types}. As shown by the figure the different types differ 
mainly in their ratio between the UV and optical flux. However, the nebular 
emission line strengths of the templates also increase from SED\,I (no 
emission) to SED\,V (numerous strong emission lines). Template~VI, which was 
constructed from SED\,V spectra with strong Ly$\alpha$ emission, shows a 
similar continuum as SED\,V.

As they were composed from spectra of different redshift the mean spectra 
cover large wavelength ranges. Since different spectral regions originate
from objects of different redshifts, it is clear that the mean spectra may 
not correspond to real galaxy spectra, if evolutionary effects are 
significant. Nevertheless, they are well suited for the purpose of deriving 
redshifts and rough object types.

\subsection{Redshift derivation}\label{xcorr}

To derive an accurate uniform set of redshifts, all spectra were (after 
conversion to a logarithmic wavelength scale) cross-correlated with their 
corresponding templates (see Simkin \cite{SIM74}; Tonry \& Davis 
\cite{TON79}). Only regions of the input spectra with good quality were 
included in the calculation. Furthermore, the continua of spectra and 
templates were removed to suppress the continuum background in the 
correlation function. 

The most likely redshift was derived using a $\chi^2$ test checking the 
correspondence of the respective spectrum (continuum inclusive) and the 
template, with the redshift being a free parameter. This procedure takes 
into account that the amplitude of cross-correlation peaks depends on object 
type, redshift and S/N.

\subsection{Final redshifts and spectral types}\label{redshifts}

For each spectrum the redshift derivation procedure was carried out for each 
template shown in Fig.~\ref{fig_sed}. Then the most probable redshift 
{\em and} spectral type were determined from the best fit. If the results 
for two templates indicated similar redshifts and fit quality, the average 
was taken. In general, the redshift differences were within the error limits 
(see below).  
 
Since low S/N or unusual SEDs can introduce errors in automatic procedures,
all results of the automatic redshift derivation were verified by visual
inspection. In this way 341 redshifts could be confirmed to be obviously 
correct (277) or very likely (about 90\,\% confidence) correct (64). These 
objects are listed in Table~\ref{tab_cat}. They represent 88\,\% of the 
primary and 23\,\% of the secondary targets. Considering the objects with 
secure redshifts only, the values are 77\,\% and 14\,\%. For the remaining 
263 objects the redshift derivation was either ambiguous or no redshift 
could be derived because of too low a S/N.  

To estimate the redshift accuracy we compared our results with those 
obtained from higher-resolution spectra available for part of our galaxies. 
For 51 brighter objects with $z < 1$, we made use of medium resolution 
FORS-600\,R spectra ($\Delta\lambda \sim 4$\,\AA{}) obtained for a different 
programme (see Ziegler et al. \cite{ZIE02}; B\"ohm et al. \cite{BOH04}). The 
comparison indicates relative redshift errors 
$\Delta z / (z + 1) \sim 4 \times 10^{-4}$ for the 150\,I spectra showing 
emission lines. For type~I spectra, which are characterised by relatively 
weak absorption features, the accuracy is about $8 \times 10^{-4}$.

To estimate the redshift uncertainties for all spectra, especially the 
high-redshift ones, the dependence of the redshift on the template used was 
taken into account. Moreover, redshifts were also derived by measuring 
spectral line positions. Taking together the different error estimates the 
typical relative uncertainties range between $4 \times 10^{-4}$ and 
$10 \times 10^{-4}$ depending on redshift regime, spectral type and S/N. 
Values for the individual redshift errors are presented in 
Table~\ref{tab_cat}.

\section{The Catalogue}\label{catalogue}

\begin{figure}
\centering 
\includegraphics[width=6cm,angle=-90]{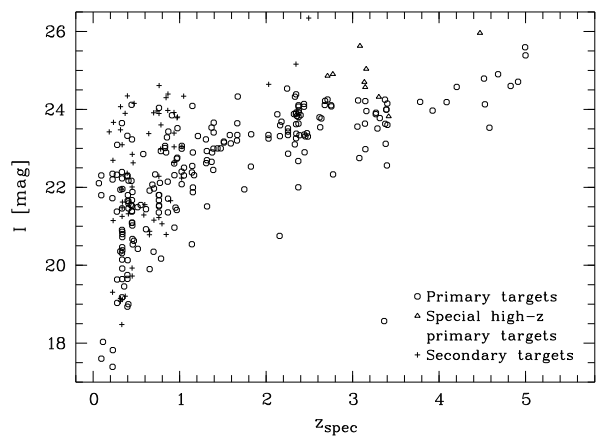}
\caption[]{$I$ magnitudes as a function of the spectroscopic redshifts of 
the extragalactic Catalogue objects. Primary and secondary targets are 
indicated by circles and crosses, respectively. In contrast to 
Fig.~\ref{fig_Iz}, the special primary objects discussed in 
Sect.~\ref{sample} are marked by triangles. The bright object with 
$z = 3.365$ is the quasar Q\,0130--260.}
\label{fig_Izs}
\end{figure}

\begin{table*}
\caption[]{Excerpt from the electronically available Catalogue of the FDF 
spectroscopic sample. Descriptions of the individual columns are given in 
the text.}
\label{tab_cat}
\centering
\begin{tabular}{c c c c c c c c c c c c c c c}
\hline
\noalign{\smallskip}
No. & RA & DEC & $B$ & $I$ & $T_{\rm exp}$ & $f / f_0$ & S/N & 
$Q_{\rm s}$ & Type & $z$ & d$z$ & $\sigma$ & $Q_{\rm z}$ & Rem. \\
& 01:$\ldots$ & $-$25:$\ldots$ & [mag] & [mag] & [min] & [\%] & (Band) & 
& & & & [\%] & & \\
\noalign{\smallskip}
\hline
\noalign{\smallskip}
4522 & 06:03.8 & 45:42 & 99.0 & 25.4 &1121 & 97 &  3.4 ($I$) & 2 & 
 V   & 4.9996 & 0.0048 &  43 & 2 & \\  
4573 & 06:04.0 & 46:55 & 25.5 & 23.9 & 281 & 31 &  3.6 ($R$) &   & 
 IV  & 0.6993 & 0.0007 &  41 & 2 & \\  
4609 & 06:04.2 & 46:45 & 25.3 & 23.4 & 135 & 47 &  4.1 ($g$) &   & 
 IV  & 0.6431 & 0.0007 &  50 & 1 & \\ 
4654 & 06:04.3 & 47:19 & 22.2 & 20.1 &  30 &  5 &  4.0 ($R$) &   & 
 IV  & 0.3989 & 0.0006 &  40 & 2 & \\ 
4657 & 06:04.3 & 43:20 & 21.8 & 19.3 & 120 &  2 &  8.8 ($g$) & 1 & 
III  & 0.2244 & 0.0005 &  36 & 1 & 600R \\  
4662 & 06:04.3 & 43:12 & 25.4 & 24.0 & 240 & 35 &  3.5 ($g$) & 2 & 
 V   & 0.7386 & 0.0007 &  50 & 2 & \\  
4667 & 06:04.3 & 47:14 & 22.6 & 20.7 &  30 & 16 &  7.5 ($R$) &   & 
 IV  & 0.4542 & 0.0006 &  32 & 1 & \\  
4682 & 06:04.4 & 46:15 & 99.0 & 23.6 & 510 & 49 &  6.1 ($g$) & 2 & 
 V   & 3.1428 & 0.0033 &  18 & 1 & \\  
4683 & 06:04.4 & 46:51 & 20.2 & 18.6 & 155 & 54 &284.2 ($g$) &   & 
VII  & 3.3650 & 0.0070 &     & 1 & QSO \\  
4684 & 06:04.4 & 48:00 & 25.7 & 23.6 &  30 & 87 &  2.7 ($g$) &   & 
 IV  & 0.7843 & 0.0007 &  52 & 2 & \\  
4685 & 06:04.4 & 43:11 & 25.0 & 23.5 & 240 & 50 &  9.4 ($g$) &   & 
 IV  & 2.6254 & 0.0025 &  16 & 1 & \\  
4686 & 06:04.4 & 45:54 & 22.7 & 21.8 & 442 & 54 & 43.8 ($g$) &   & 
 IV  & 0.0943 & 0.0004 &  15 & 1 & \\ 
4691 & 06:04.4 & 48:34 & 25.8 & 24.3 & 550 & 66 & 13.1 ($g$) &   &
V/VI & 3.3036 & 0.0043 &  67 & 1 & LAB \\ 
4701 & 06:04.5 & 42:52 & 29.4 & 24.7 & 225 & 60 &  1.4 ($I$) &   &
 V   & 4.9132 & 0.0047 &  82 & 2 & \\
4729 & 06:04.6 & 47:36 & 22.3 & 21.1 & 120 & 51 & 34.2 ($g$) &   & 
VIII & 0      &        &     & 1 & G \\  
4745 & 06:04.6 & 44:03 & 25.2 & 23.8 & 434 & 67 & 10.2 ($g$) &   & 
 V   & 2.6173 & 0.0029 &  17 & 1 & \\  
4752 & 06:04.7 & 46:54 & 26.5 & 99.0 & 473 & 29 &  3.2 ($g$) & 2 & 
V/VI & 3.3802 & 0.0044 & 119 & 1 & LAB \\
4795 & 06:04.8 & 47:14 & 24.4 & 23.3 & 570 & 55 & 15.9 ($g$) &   & 
 IV  & 2.1593 & 0.0022 &  14 & 1 & \\
4871 & 06:05.1 & 46:04 & 24.9 & 23.4 & 490 & 58 & 15.4 ($g$) &   & 
 IV  & 2.4724 & 0.0024 &  13 & 1 & \\
4882 & 06:05.1 & 48:38 & 21.3 & 19.0 & 120 & 31 & 75.0 ($R$) &   & 
 II  & 0.2777 & 0.0006 &  11 & 1 & \\  
4910 & 06:05.2 & 46:04 & 25.5 & 24.3 & 120 & 40 &  2.1 ($g$) &   & 
 V   & 0.8437 & 0.0007 &  65 & 2 & \\  
4954 & 06:05.4 & 46:20 & 25.8 & 23.8 & 160 & 37 &  2.1 ($R$) &   & 
 IV  & 0.9703 & 0.0008 &  55 & 2 & \\ 
4993 & 06:05.5 & 45:56 & 26.5 & 21.8 & 245 & 50 & 11.7 ($I$) &   & 
 I   & 0.7632 & 0.0014 &  24 & 1 & \\ 
4996 & 06:05.5 & 46:28 & 24.4 & 23.3 & 281 & 56 & 14.6 ($g$) &   & 
 IV  & 2.0278 & 0.0021 &  15 & 1 & \\ 
\noalign{\smallskip}
\hline
\end{tabular}
\end{table*}

The basic properties of all 341 objects with certain or probable redshift 
are listed in Table~\ref{tab_cat} (in the following referred to as 
`the Catalogue'), which is available in electronic form only, except for a 
sample page presented in Table~\ref{tab_cat}. The individual columns of the 
Catalogue have the following content: 
\begin{description}
\item[No.:] Object number according to Heidt et al. (\cite{HEI03b}). For a 
few objects not listed in the photometric catalogue (mainly because no 
photometry could be obtained due to crowding) new numbers, starting with 
9001, have been assigned.
\item[RA:] Right ascension (equinox: 2000) in hours, minutes and seconds. 
\item[DEC:] Declination (equinox: 2000) in degrees, arcminutes and 
arcseconds.
\item[B:] Total apparent $B$ magnitude (Vega system, see Heidt et al. 
\cite{HEI03b}). Non-detec\-tions in $B$ are marked by the value $99.0$.
\item[I:] Total apparent $I$ magnitude (as above). 
\item[$T_{\rm exp}$:] Total exposure time in minutes.
\item[$f / f_0$:] Ratio between the flux which passed through the slit and
the actual object flux in \%. Low $f / f_0$ usually correspond to large 
object extensions. Typical values for point-like objects are around 70\,\%. 
Large values ($> 80$\,\%) can be caused by systematic spectral errors and/or 
inaccurate photometry due to very low fluxes or object crowding.
\item[S/N:] Average signal-to-noise ratio per resolution element in the 
filter band given in parentheses ($B$, $g$, $R$ or $I$). In each case the 
band with the highest S/N was selected. The S/N as a function of wavelength 
was calculated by dividing the object spectrum by its error function.
\item[$Q_{\rm s}$:] Flag indicating systematic errors in the spectrum.   
$Q_{\rm s} = 1$ refers to distorted spectra, $Q_{\rm s} = 2$ to local 
defects.
\item[Type:] Object type as defined in Sect.~\ref{types} and 
Fig.~\ref{fig_sed}. The classes I to VI represent galaxies, VII indicates 
QSOs, and VIII refers to stars.        
\item[$z$:] Spectroscopic redshift.
\item[d$z$:] Mean error of the redshift.
\item[$\sigma$:] Relative rms deviation between spectrum and the optimal 
template in \% of the average spectral flux.
\item[$Q_{\rm z}$:] Quality of the redshift. $Q_{\rm z} = 1$ indicates 
objects with secure redshifts and $Q_{\rm z} = 2$ with probable redshifts 
($90\,\%$ confidence level).
\item[Rem.:] Further information on the object. For stars a rough spectral 
type is given. Quasars and strong Ly$\alpha$ emission galaxies are indicated 
by the entries `QSO' and `LAB' (Ly$\alpha$ bright, i.e. Ly$\alpha$ emission 
EW $\ge 20$\,\AA{}), respectively. `600R' indicates galaxies whose redshift 
and object type were verified by means of the spectroscopic data of the 
medium resolution spectra (see Sect.~\ref{redshifts}).     
\end{description}

To illustrate the sample properties Fig.~\ref{fig_Izs} shows the $I$ 
magnitudes of the extragalactic Catalogue objects as a function of 
spectroscopic redshift. Primary (236) and secondary objects (63) are 
indicated. High-redshift primary targets, which were selected in a special 
way (12 objects, see Sect.~\ref{sample}), are marked by triangles to 
distinguish them from the $I$ magnitude selected primary targets, which are 
indicated by circles.

\section{Basic properties of the spectra}\label{propspec}

\begin{figure*}
\centering
\includegraphics[width=6cm,angle=-90]{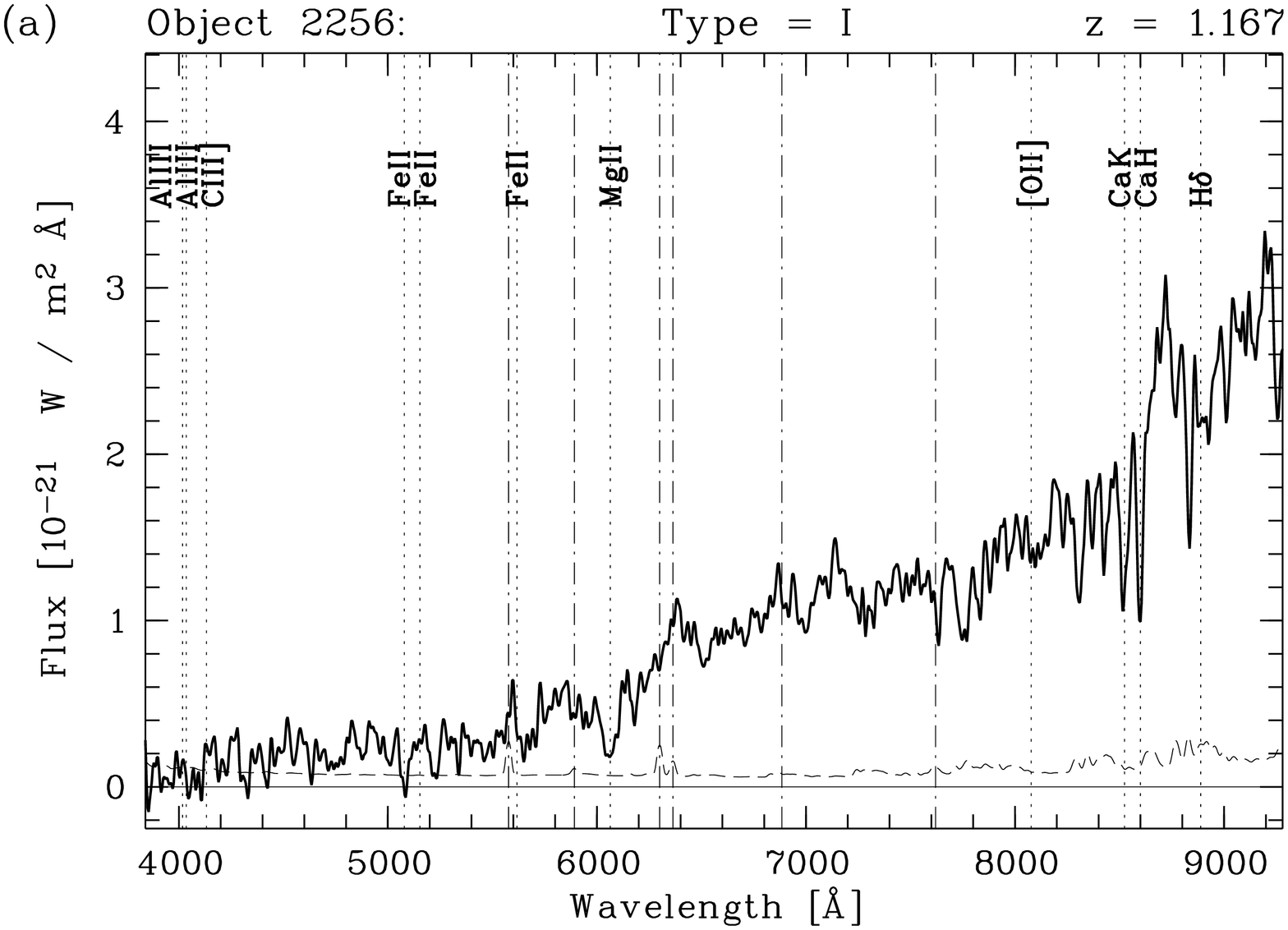}
\includegraphics[width=6cm,angle=-90]{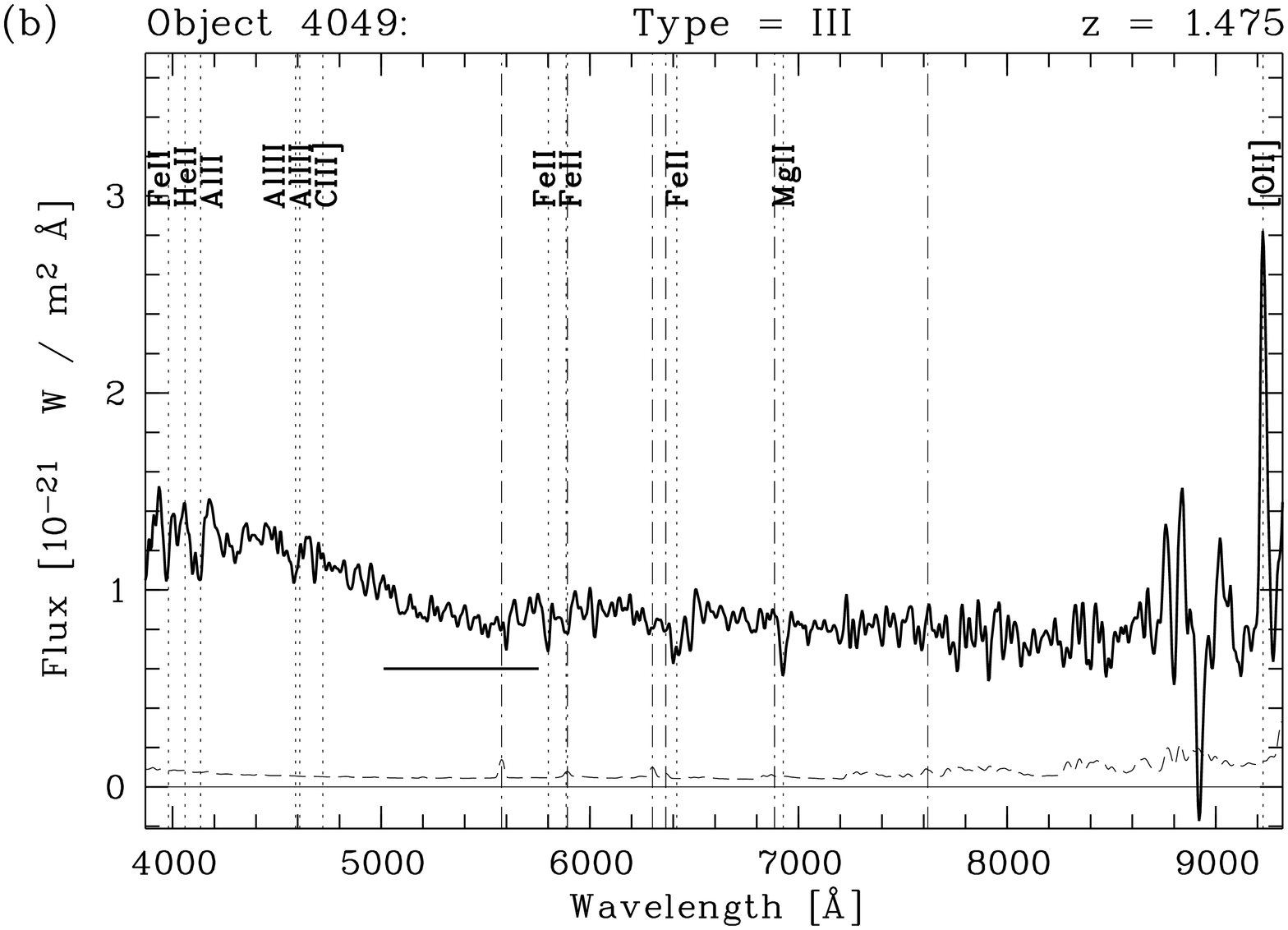}
\includegraphics[width=6cm,angle=-90]{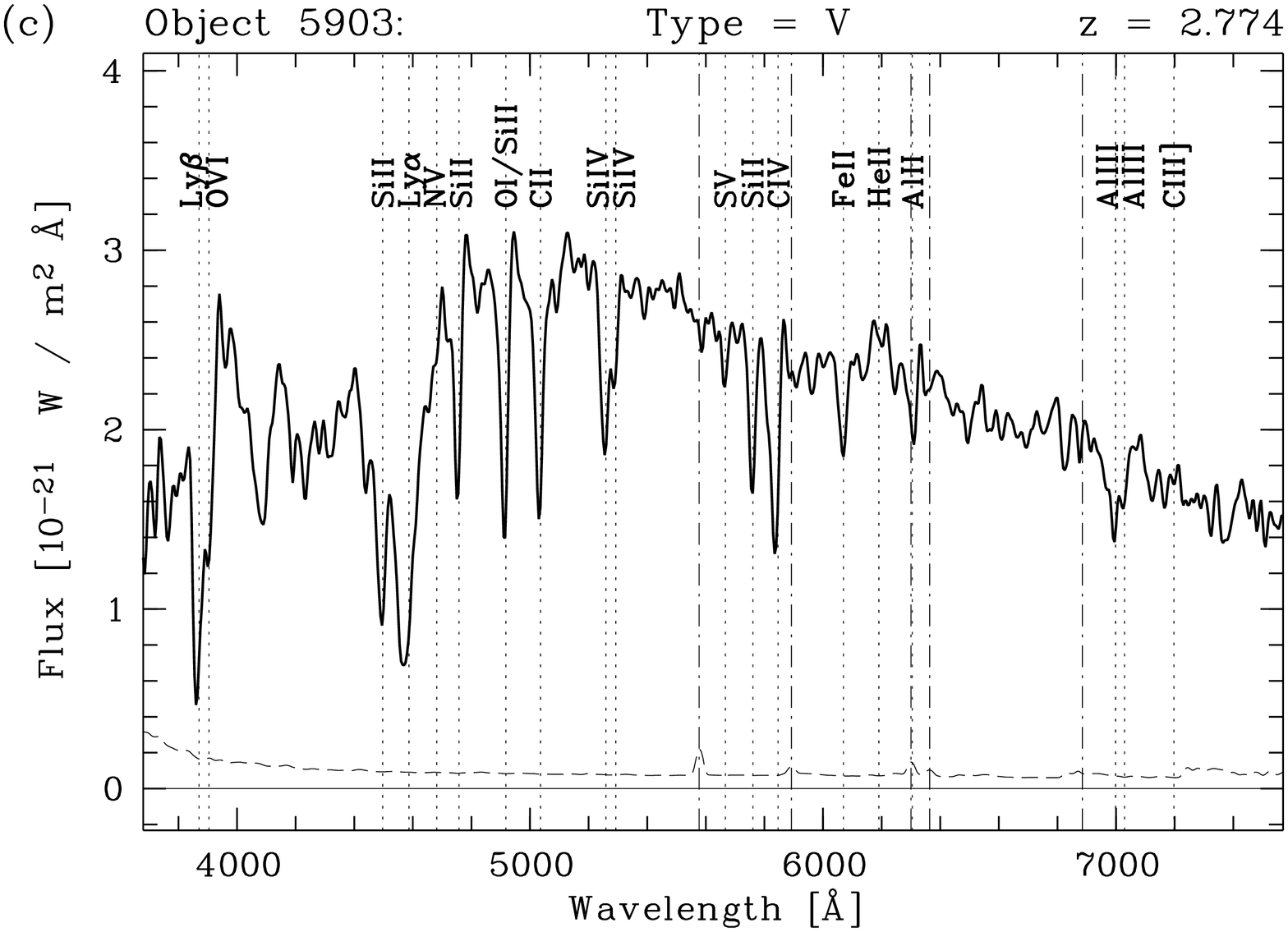}
\includegraphics[width=6cm,angle=-90]{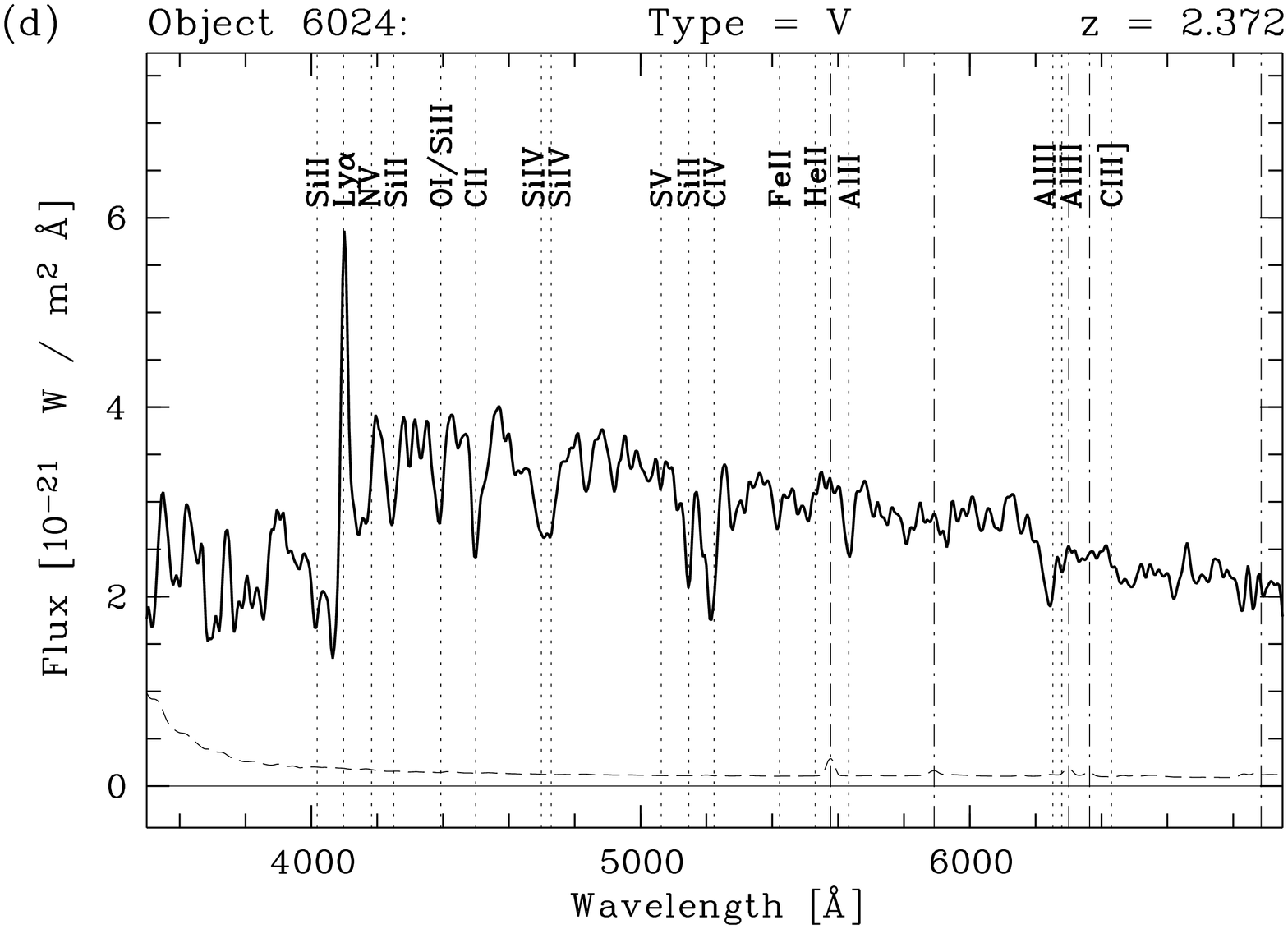}
\includegraphics[width=6cm,angle=-90]{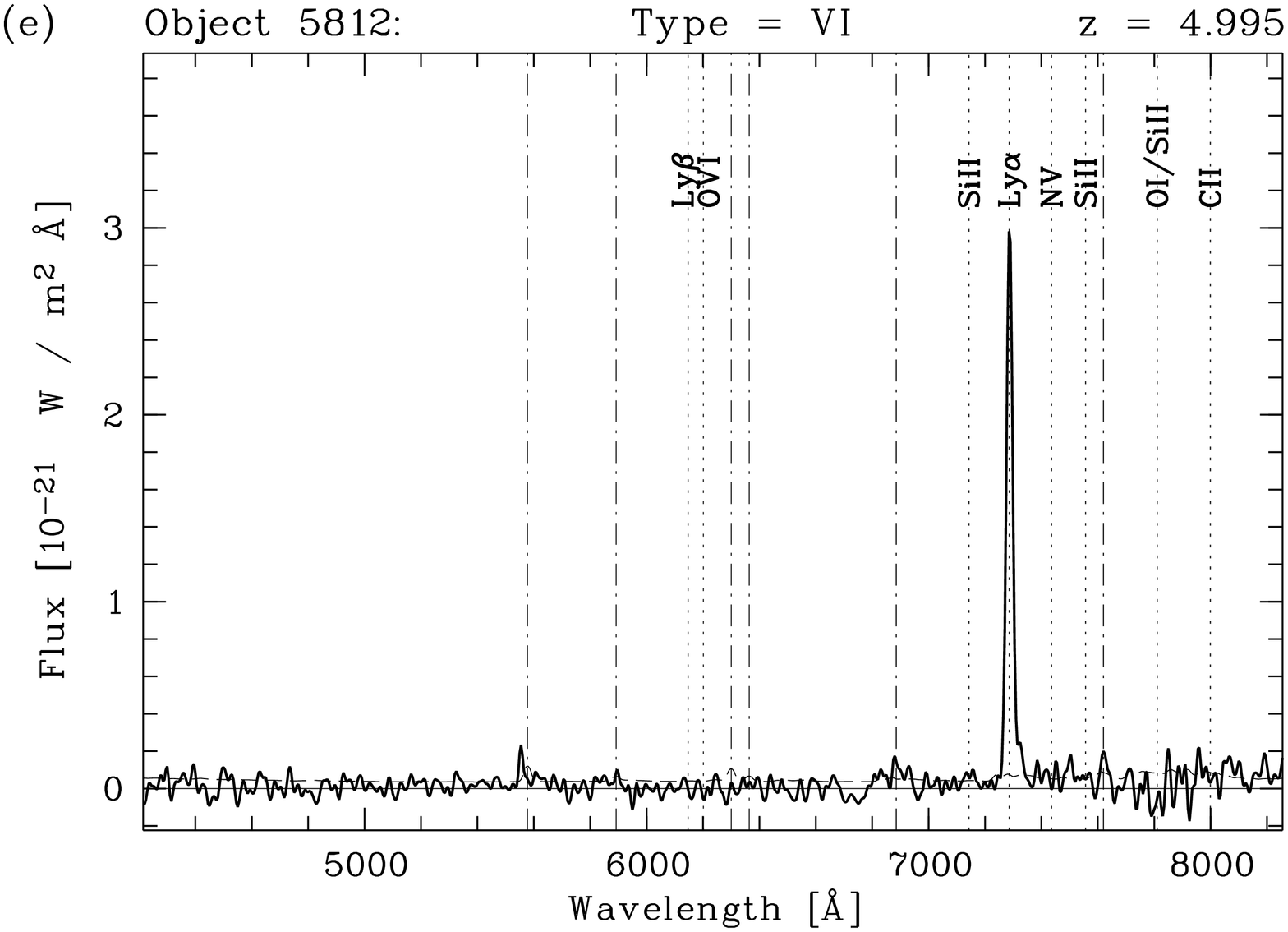}
\includegraphics[width=6cm,angle=-90]{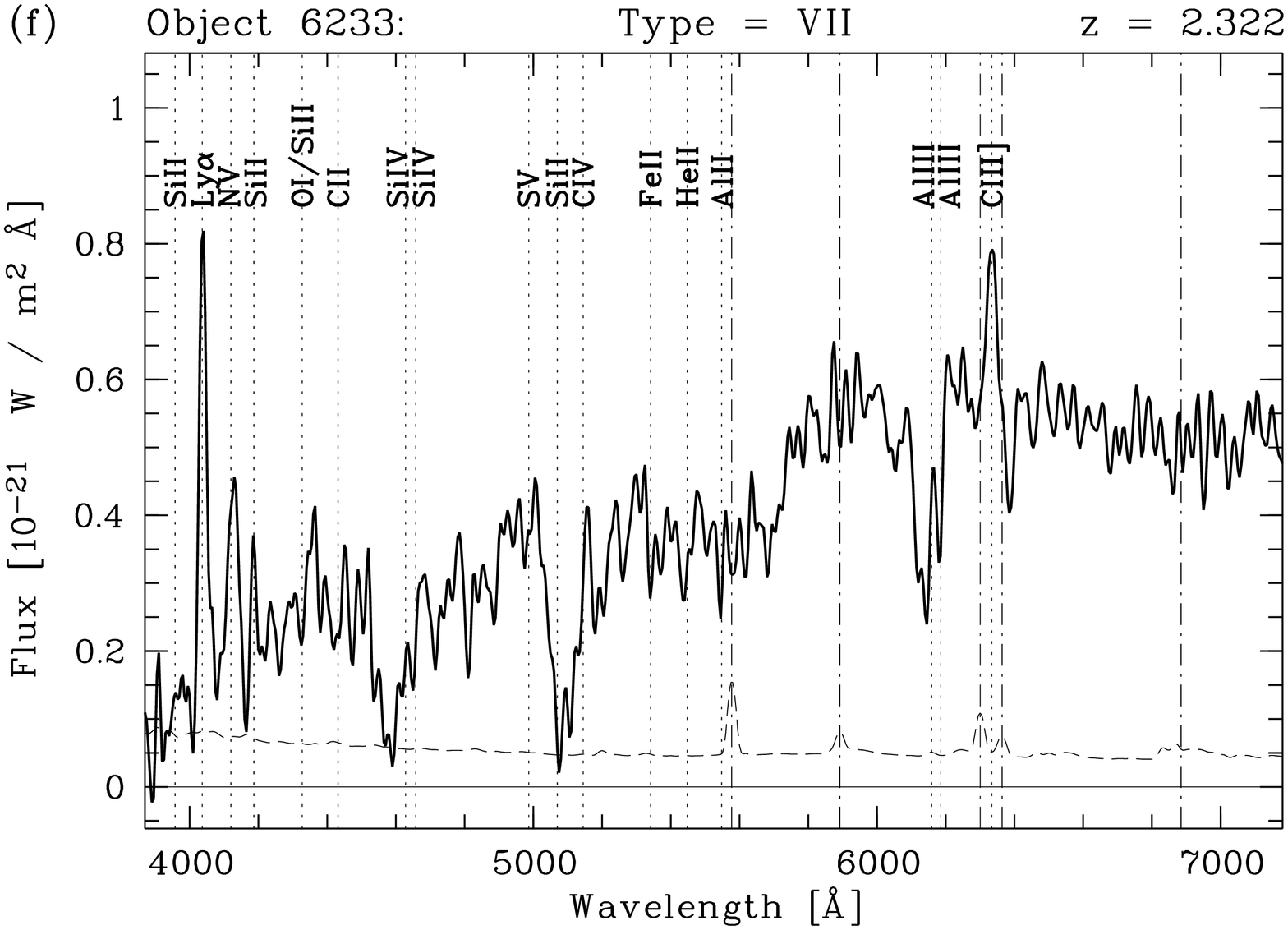}
\caption[]{Examples of the FDF spectra. The positions of some prominent 
spectral lines are marked by vertical dotted lines. In panel (b) the 
position of the UV absorption feature at 2200\,\AA{} is indicated by a 
horizontal bar. Vertical dash-dotted lines indicate the positions where
terrestrial atmospheric features ([O\,I]\,$\lambda\,5577$, Na\,I\,D, 
[O\,I]\,$\lambda\lambda\,6300,6364$, B~band, and A~band) were removed. The 
noise level is indicated by a dashed line. At the top of each diagram the 
object number, the spectral type, and the redshift are given.}
\label{fig_galspec} 
\end{figure*}

In addition to the catalogue the flux-calibrated individual object spectra 
with reliable redshifts and no strong systematic errors are also available 
as an electronic atlas. In the following we describe some properties of 
these spectra.

\subsection{Stars}\label{stars}

The FDF spectroscopic sample contains 42 stars of spectral types G to L. Most 
frequent are K stars (18). Five stars have spectral types later than M5.
None of the stars is brighter than 17\,mag in $I$. Since the FDF is located 
close to the South Galactic Pole, most of the observed stars must belong to 
the halo or the thick disk.

\subsection{Galaxies}\label{galaxies}

The spectra of the 158 galaxies with $z < 1$ show a great variety of spectral
types ranging from ellipticals to extreme starburst galaxies. 
       
Since early-type galaxies are intrinsically red, it is difficult to find such 
objects with high redshifts in optical surveys. For $z > 1.3$ the 4000\,\AA{} 
break is shifted beyond our spectral window. This readily explains why no 
type~I galaxy beyond $z = 1.17$ (FDF-2256, see Fig.~\ref{fig_galspec}(a)) 
could reliably be identified in the FDF, so far, and why only six candidates 
of spectral type I or II with $z \ge 1$ were detected.   

Some of the intermediate redshift galaxies show a depression at about 
2200\,\AA{} (rest frame). As an example FDF-4049 is presented in 
Fig.~\ref{fig_galspec}(b). The 2200\,\AA{} feature of this galaxy 
corresponds to $E_{\rm B - V} \approx 0.15$\,mag if a Galactic extinction 
law is assumed. The figure seems to illustrate that the well-known broad dust 
absorption feature, which is thought to be caused by small carbon-rich 
particles (see Cardelli et al. \cite{CAR89}), is also present in at least 
part of the intermediate-redshift galaxies.
  
In the rest-frame wavelength range between 2000 and 3700\,\AA{} galaxy 
spectra are poor in prominent spectral lines. This complicates the 
derivation of reliable redshifts in the range between $1.5$ and 2. Therefore, 
in this redshift range secure redshifts could be derived for two objects 
only.

For $z \ge 2$ we observe only starburst galaxies. Their spectra are dominated 
by the Ly$\alpha$ line (with profiles ranging from strong absorption to 
strong emission), the absorptions of the Ly$\alpha$ forest short-ward of 
1216\,\AA{}, and the spectral break of the Lyman limit at 912\,\AA{}.  
Furthermore, the spectral region red-ward from Ly$\alpha$ is marked by 
numerous lines, usually showing absorption profiles, and often of 
interstellar origin (e.g. lines of Si\,II and C\,II). Except for Ly$\alpha$ 
significant emission could be detected in a few cases only, either in the 
form of a P\,Cygni profile (e.g. C\,IV\,$\lambda\lambda\,1548,1550$) or as 
pure emission (e.g. He\,II\,$\lambda\,1640$ and 
C\,III]\,$\lambda\lambda\,1907,1909$). All these features are visible in the 
spectra of FDF-5903 (Fig.~\ref{fig_galspec}(c)) and FDF-6024 
(Fig.~\ref{fig_galspec}(d)). 
 
Four (44\,\%) of the nine galaxies with $z \gtrsim 4.5$ were found to be 
Ly$\alpha$ bright (type~VI). This percentage is comparable to the redshift 
range $3 < z < 4$ where about 37\,\% are LABs. Up to now, the objects with 
the highest spectroscopic redshifts in the FDF are the galaxies FDF-4522 and 
FDF-5812 (Fig.~\ref{fig_galspec}(e)) both having $z \approx 5.0$.

\subsection{Quasars}\label{quasars}

Most of the eight quasars at redshifts between $0.865$ and $3.365$ identified 
in the FDF show normal QSO spectra dominated by broad emission lines of 
Ly$\alpha$, N\,V, Si\,IV, C\,IV, C\,III] and/or Mg\,II. The only exception is 
FDF-6233 at $z = 2.32$ (see Fig.~\ref{fig_galspec}(f)) which belongs to the 
rare class of broad absorption line quasars (BAL, e.g. Menou et al. 
\cite{MEN01}). Its spectrum is dominated by strong, broad, blue-shifted 
absorption troughs of Si\,IV, C\,IV and Al\,III indicating outflows of 
several 1000\,km/s. Ly$\alpha$, N\,V, and C\,III] show relatively narrow 
emission components. In contrast to the other quasars the continuum of 
FDF-6233 decreases rapidly short-ward of 1750\,\AA{} (rest frame).

\section{Distribution of spectroscopic redshifts}\label{distribution}

\begin{figure}
\centering 
\includegraphics[width=8.8cm]{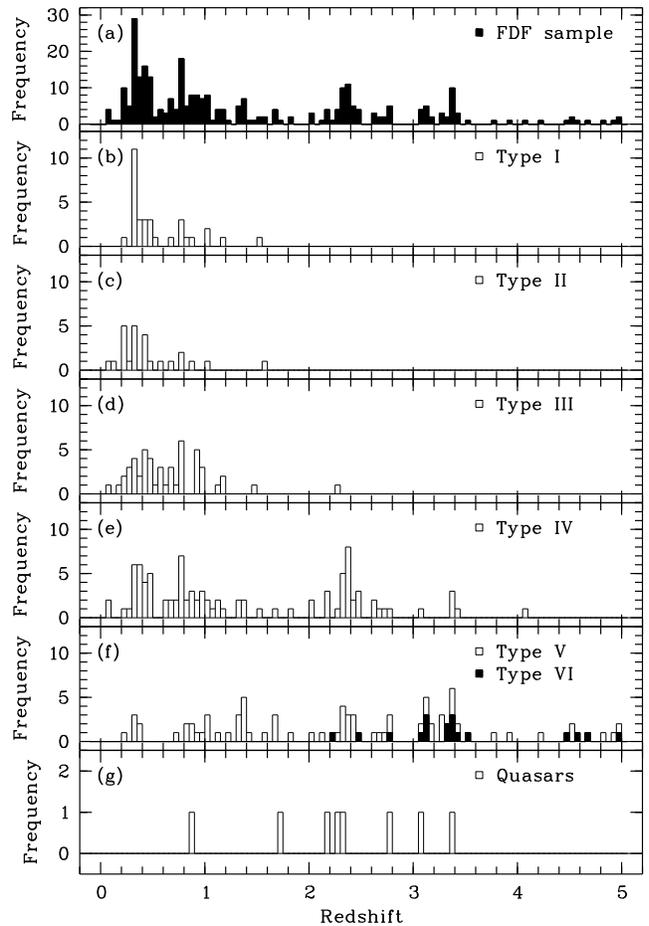}
\caption[]{Redshift distribution of the FDF spectroscopic sample for (a) the
complete Catalogue (see Sect.~\ref{catalogue}), (b) -- (f) galaxies of 
different object type (see Fig.~\ref{fig_sed}) and (g) the quasars. In 
diagram (f) galaxies of type V (white bars) as well as type VI (black bars) 
are plotted. The redshift resolution is $\Delta z = 0.05$.}
\label{fig_tz}
\end{figure}

\begin{figure}
\centering 
\includegraphics[width=8.8cm,clip=true]{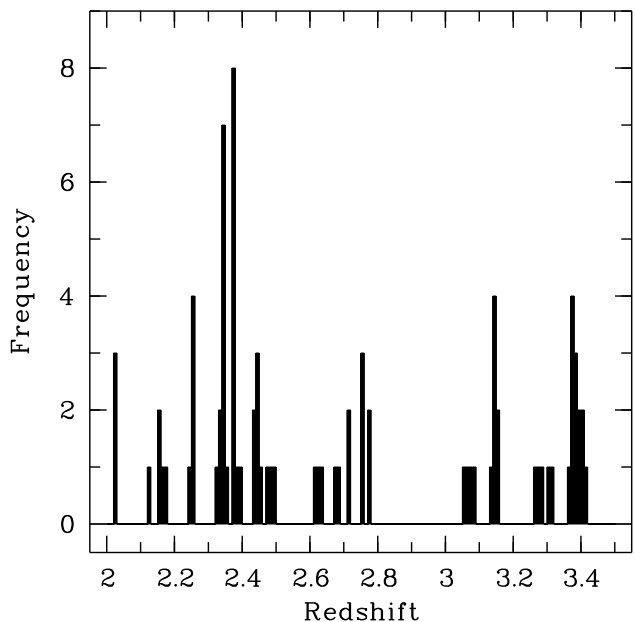}
\caption[]{Redshift distribution of the FDF spectroscopic sample in the 
range $2 < z < 3.5$ plotted with a redshift resolution of $\Delta z = 0.01$.}
\label{fig_thz}
\end{figure}

\begin{figure}
\centering 
\includegraphics[width=8.8cm,clip=true]{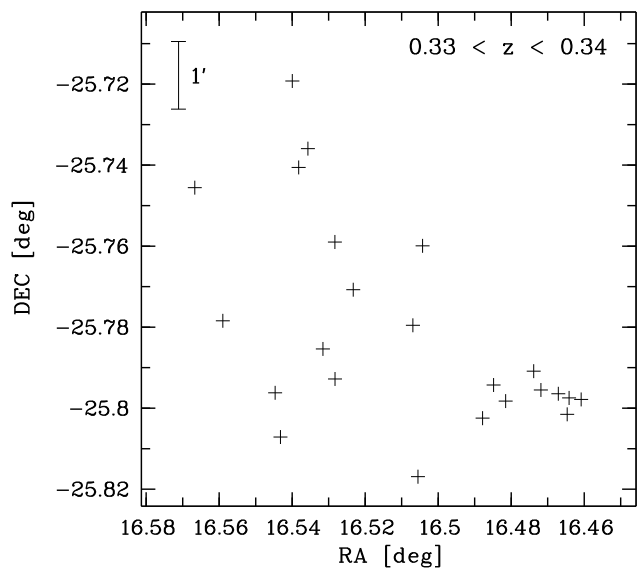}
\caption[]{Distribution of galaxies within the range $0.33 < z < 0.34$ in the 
FDF.}
\label{fig_clust}
\end{figure}

\begin{figure}
\centering 
\includegraphics[width=8.8cm,clip=true]{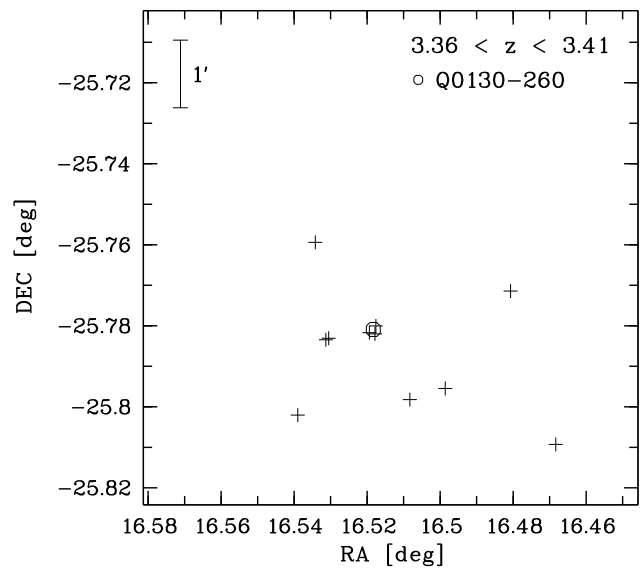}
\caption[]{Distribution of galaxies within the range $3.36 < z < 3.41$ in the 
FDF. The position of the quasar Q\,0130--260 is indicated by a circle. 
The three objects close to the quasar are strong Ly$\alpha$ emitters, which 
were selected using narrow-band images (see Sect.~\ref{sample}).}
\label{fig_qsoclust}
\end{figure}

Fig.~\ref{fig_tz}(a) shows the spectroscopic-redshift distribution of the 
galaxies and QSOs with a resolution of $\Delta z = 0.05$. Significant local 
maxima are evident at $z = 0.325$ (29 objects), $z = 0.775$ (18), $z = 2.375$ 
(11) and $z = 3.375$ (10). In part this clustering of redshifts had already 
been noticed in the distribution of the photometric redshifts. In some cases 
the overdensities extend over more than two bins. For instance, the three 
adjacent bins comprising the redshift range between $2.30$ and $2.45$ contain 
26 objects which represent about 27\,\% of the $z \ge 2$ sample. On the other 
hand, the peak around $z = 0.325$ is mainly produced by a redshift interval 
ranging from $0.331$ to $0.339$ ($\langle z \rangle = 0.335$, 
$\sigma_z = 0.002$) which is smaller than the bin size but which includes 23 
galaxies. As indicated in Fig.~\ref{fig_thz} such fine structure of the 
redshift distribution is also present at high redshifts. An example is the 
narrow redshift range between $2.372$ and $2.378$ 
($\langle z \rangle = 2.374$, $\sigma_z = 0.002$) which includes eight 
galaxies, or about 73\,\% of the objects within the standard bin around 
$z = 2.375$ (see above). As pointed out by Frank et al. (\cite{FRA03}) the 
high-redshift galaxy overdensities correlate well with QSO metal absorption 
line systems in the direction of the FDF. Besides the striking peaks, there 
are also indications of marked gaps in the redshift distribution. At low 
redshifts this is particularly evident around $z = 0.525$ where the redshift 
distribution reaches a local minimum. At high redshifts a lack of galaxies 
exists between $2.80$ and $3.05$, where no objects could be detected, so far. 
The relatively low number of objects between $1.5$ and $2.0$ is a 
consequence of the difficulty to identify galaxies in this redshift range 
from visual spectra (see Sect.~\ref{galaxies}).

Since the spectroscopic sample is representative for the most luminous 
objects at each redshift (see Sect.~\ref{sample}), the observed 
distribution can (apart from the redshift range $1.5 < z < 2.0$) be assumed
to be representative of the real distribution for the absolutely bright 
objects in the direction of the FDF. On the other hand, we have almost no 
information on the distribution of the intrinsically faint $z > 2$ objects 
($M_{\rm B} > -20$\,mag), as such objects are rare in our sample.

Since the first photometric observations of the FDF a (poor) cluster is known 
to be located in the southwestern corner of the field (see Heidt et al. 
\cite{HEI03b}, Ziegler et al. \cite{ZIE04}). This cluster contributes to the 
peak of the FDF redshift distribution at $z \approx 0.335$. As indicated by 
Fig.~\ref{fig_clust} the other objects at this redshift are not randomly 
distributed either, as no single object of this redshift is found in the 
northwestern corner of the FDF. For the other peaks no significant deviation 
from a random distribution over the area of the FDF could be detected. This 
negative result may in part be due to the insufficient number of objects. 
But, since the angular size of the FDF covers just about the size of a large 
galaxy cluster at high redshift no major angular clustering effects were 
expected. Nevertheless, some indications for clustering on small scales were 
found. For instance, the galaxies between $z = 3.36$ and $3.41$ populate the 
southern part of the FDF only (Fig.~\ref{fig_qsoclust}). Furthermore, three 
Ly$\alpha$ emitters of this redshift were found less than $5''$ from the 
bright quasar Q\,0130--260. Moreover, several close pairs of galaxies with 
similar redshifts were discovered. In one case a dense group at 
$z \approx 2.347$, comprising four members (FDF-5135,\,5165,\,5167,\,5190), 
was identified. The mutual distances of these objects are $< 3''$.      

In general, the FDF spectroscopic data, indicating predominantly moderate 
clustering of galaxies up to high redshifts, which can be attributed to 
large-scale structure, agree well with earlier results of, e.g., Adelberger 
et al. (\cite{ADE98}), Steidel et al. (\cite{STE98}), or Cohen et al. 
(\cite{COH00}).   
   
\begin{table}
\caption[]{Distribution of objects with reliable redshifts in the FDF 
spectroscopic sample. The table indicates for each type the total number of 
objects $N$, the median redshift $\overline{z}$, the average redshift 
$\langle{}z\rangle$ and its variance $\sigma_z$.}
\label{tab_types}
\centering
\begin{tabular}{c c c c c}
\hline
\noalign{\smallskip}
Type & $N$ & $\overline{z}$ & $\langle{}z\rangle$ & $\sigma_z$ \\
\noalign{\smallskip}
\hline
\noalign{\smallskip}
I             &  32 & 0.4 & 0.6 & 0.3 \\ 
II            &  25 & 0.3 & 0.5 & 0.3 \\
III           &  50 & 0.6 & 0.7 & 0.4 \\
IV            &  96 & 1.0 & 1.4 & 1.0 \\
V + VI        &  88 & 2.4 & 2.4 & 1.3 \\
QSOs          &   8 & 2.3 & 2.3 & 0.8 \\
\noalign{\smallskip}
\hline
\noalign{\smallskip}
Extragalactic & 299 & 0.9 & 1.4 & 1.2 \\
objects       &     &     &     &     \\
Stars         &  42 & --  & --  & --  \\
\noalign{\smallskip}
\hline
\end{tabular}
\end{table}

The redshift distribution of the different object types defined in 
Sect.~\ref{types} and \ref{templates} is shown in 
Fig.~\ref{fig_tz}(b)--(g). The type-dependent median and mean redshifts are
presented in Table~\ref{tab_types}. For the reasons outlined above the types 
I and II are concentrated at low redshifts (median values $0.4$ and $0.3$). 
Furthermore, type~I shows a strong overdensity at $z \approx 0.335$ where 
11 of 32 early-type galaxies are located. This obviously reflects the 
predominance of E and S0 galaxies in the cluster at this redshift. The change 
of the apparent composition of object types with redshift due to the 
redshift-dependent selection bias is particularly obvious for class~V/VI: At 
low redshifts this type plays a minor role (7\,\% of the observed objects at 
$z < 0.5$), but at high redshifts type~V/VI becomes the dominating class. At 
$2 \le z < 3$ the extreme starburst galaxies contribute 38\,\% to all 
identified objects, at $3 \le z < 4$ they present 78\,\% and at $z \ge 4$ the 
fraction increases to 91\,\% (10 of 11 objects).

\section{Exploring evolutionary effects}\label{specsample}

A detailed discussion of the whole FDF spectroscopic sample is beyond the 
scope of this paper. Instead, we will focus on some properties of the 
high-redshift objects. As average spectra of spectroscopic subsamples 
selected by redshift and/or spectral type are known to be well suited to 
investigate the characteristic properties of galaxy populations, the 
following discussion is based mainly upon such mean spectra. An analysis of 
individual object spectra can be found in, e.g., Mehlert et al. 
(\cite{MEH02}) and Tapken et al. (\cite{TAP04}).

\subsection{The calculation of composite spectra}\label{calcmean}

Composite spectra of FDF subsamples were calculated using similar procedures 
as used by \cite{SHA03} to facilitate a comparison with their spectroscopic 
sample (see Sect.~\ref{comp_lbg}). In detail, the flux-calibrated, co-added 
spectra (see Sect.~\ref{reduction}) were shifted to the vacuum rest frame, 
rebinned to a scale of 1\,\AA{} per pixel, scaled to a common mode in the 
wavelength range between 1250 and 1500\,\AA{}, and finally averaged 
unweighted to get a representative average. 

The averaging of spectra showing different continuum levels may lead to a 
systematic change of the mean equivalent width of spectral lines. A 
comparison with mean spectra based on the alternative method described in 
Sect.~\ref{templates} showed that for Ly$\alpha$ an overestimate of the true 
EW of the order 10\,\% can occur. For other lines no significant effect 
could be detected. The dependence of the Ly$\alpha$ strength on the 
normalisation method must be kept in mind when analysing the results 
described in the following (see Sect.~\ref{meanspec_lya}). 
       
In contrast to \cite{SHA03}, we did not exclude extreme flux values during 
the averaging. Instead, our averaging procedure omitted wavelength ranges 
affected by significant sky line residuals.

\subsection{The overall mean spectrum of the high-redshift sample}
\label{meanspec_all}

\begin{figure*}
\centering 
\includegraphics[width=17.8cm,clip=true]{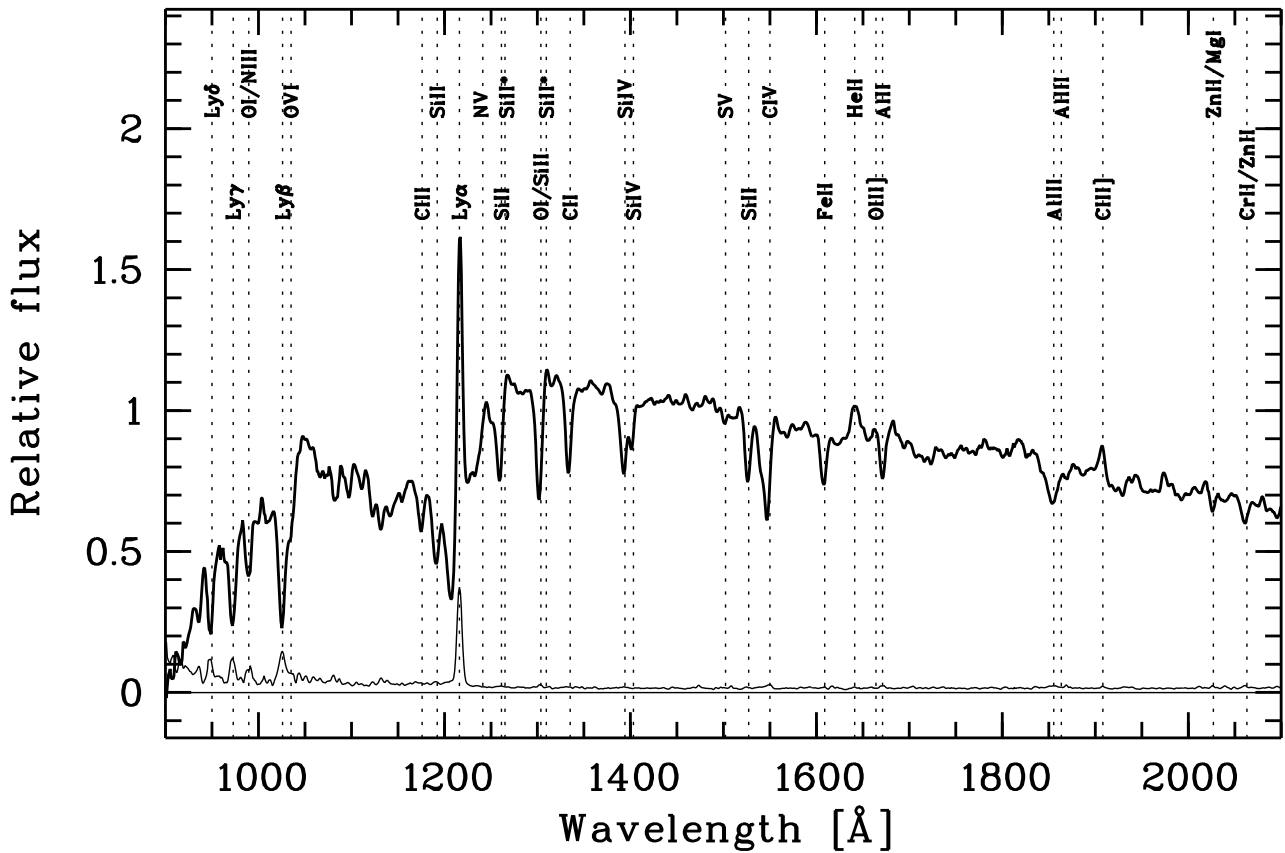}
\caption[]{Composite spectrum of 64 FDF galaxy spectra with S/N~$\ge 4$ in 
the range $ 2 < z < 4$ (thick solid line). The positions of prominent UV 
lines are indicated. The thin solid line at the bottom of the spectrum gives
the mean error of the average spectrum as derived from the scatter of the 
individual spectra. Outside strong spectral features this error 
agrees with the value expected from the observational errors. In the case of 
some strong features such as Ly$\alpha$ the mean error is dominated by the 
physical variance in the individual spectra.}
\label{fig_meanall}
\end{figure*}

\begin{figure}
\centering 
\includegraphics[width=8.8cm,clip=true]{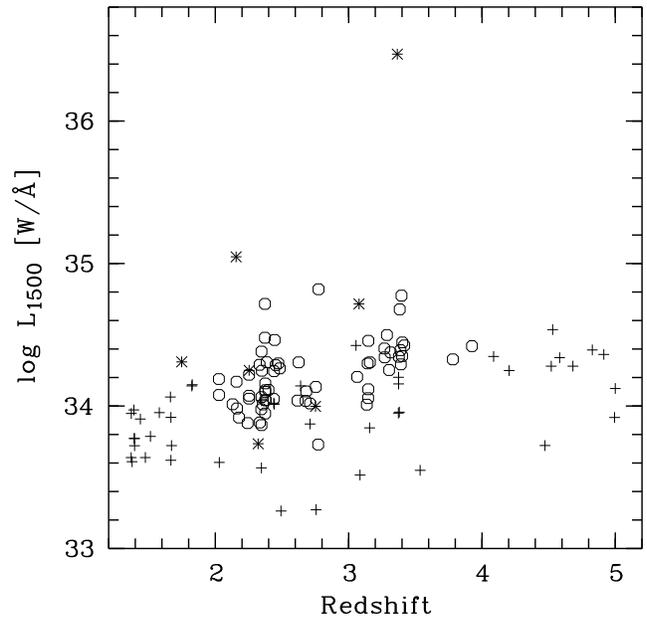}
\caption[]{Luminosities at 1500\,\AA{} of the $z > 1.35$ FDF galaxies as a 
function of redshift. To derive the luminosity the spectral flux was measured 
in the wavelength interval 1480 to 1520\,\AA{}. Galaxies used for the 
calculation of the composite spectra are indicated by circles. The remaining 
high-redshift objects are marked by crosses apart from QSOs, which are 
indicated by asterisks.}
\label{fig_luvz}
\end{figure}

In Fig.~\ref{fig_meanall} we present a composite of 64 galaxy spectra with
$2 < z < 4$ and a continuum S/N~$\ge 4$. This subsample of photometrically
selected objects ($I \lesssim 24.5$\,mag) represents about 91\,\% of all 
galaxies with secure redshift in the range $2 < z < 4$. All objects are 
primary targets (see Sect.~\ref{sample}). Three $z > 3$ galaxies 
(FDF-7452,\,7644,\,8215) were initially included in the observations because 
of suspected Ly$\alpha$ emission. One $z > 3$ object (FDF-4691) was 
originally a secondary target showing strong Ly$\alpha$ emission. One $z < 3$ 
object (FDF-9011) was included due to its location close to another sample 
galaxy. The individual spectra are representative of the integrated light of
the entire galaxies, since the slit losses are not more than twice those for 
point-like objects (see Sect.~\ref{catalogue}). The mean redshift of this 
subsample is $\langle z \rangle = 2.72$ ($\sigma_z = 0.49$). The UV 
luminosities at 1500\,\AA{} of the selected galaxies (derived from the flux 
in the interval 1480 to 1520\,\AA{}) are indicated by circles in 
Fig.~\ref{fig_luvz}. $\langle \log L_{1500} [{\rm W/\AA}] \rangle$ amounts 
to $34.22$ ($\sigma_L = 0.22$).  

The composite spectrum reaches a continuum S/N $\approx 60$, which allows a
comparison with similar high-S/N high-redshift spectra from the literature 
(e.g. Pettini et al. \cite{PET00}; \cite{SHA03}). While the general 
character of all these spectra is similar, differences in spectral details, 
such as the Ly$\alpha$ strength and the continuum slope, are evident. We 
discuss these differences in Sect.~\ref{comp_lbg}.

\subsection{The redshift dependence of spectral properties}\label{meanspec_z}

\begin{figure*}
\centering 
\includegraphics[width=17.8cm,clip=true]{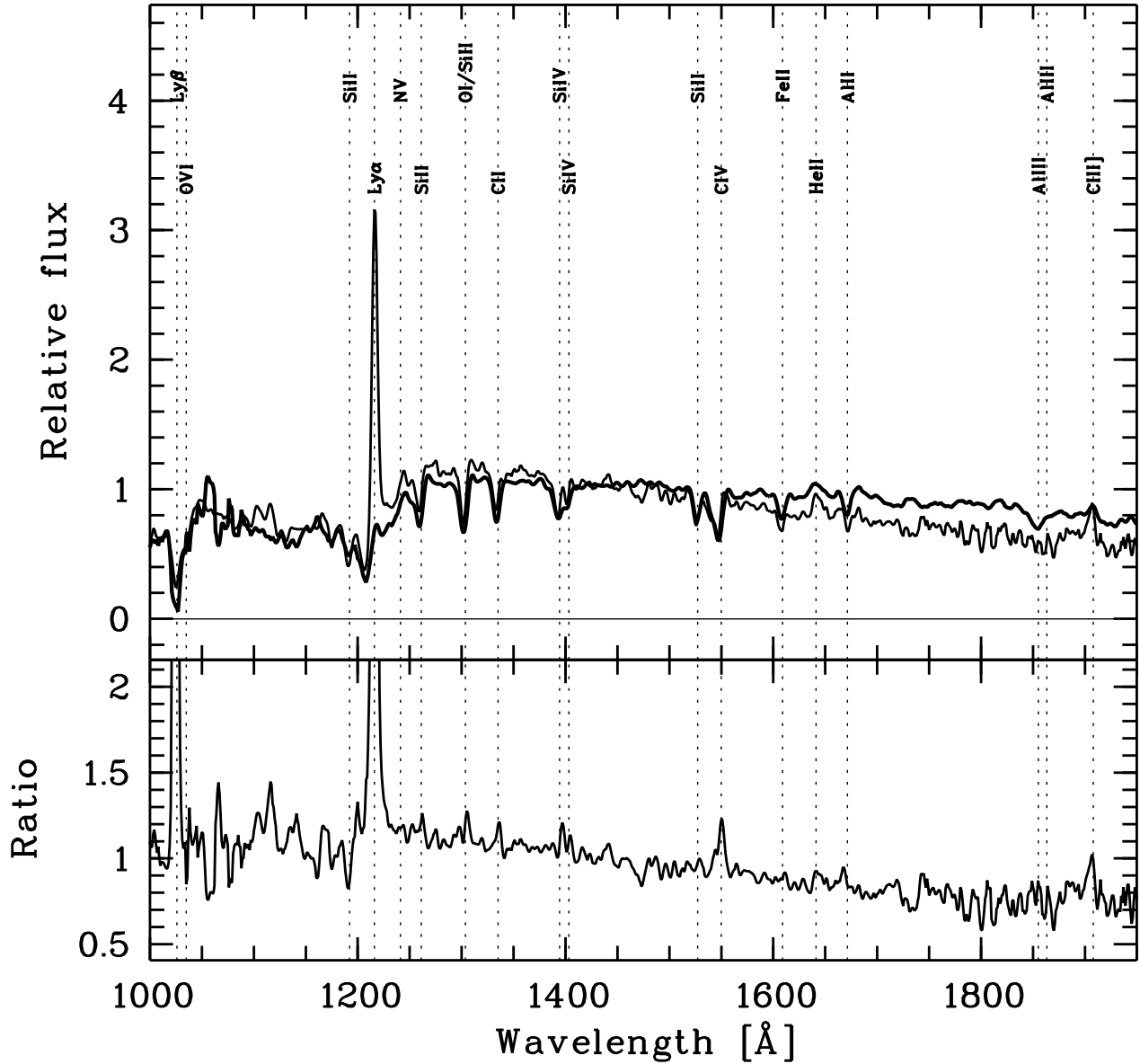}
\caption[]{Comparison of FDF composite spectra covering the redshift 
intervals $2 < z < 3$ (thick line) and $3 < z < 4$ (thin line), 
respectively. The lower panel gives the ratio of the $3 < z < 4$ and the
$2 < z < 3$ composite spectrum. The positions of prominent lines are 
indicated.}
\label{fig_mean_z}
\end{figure*}

\begin{table}
\caption[]{Properties of the FDF composite spectra representative of the 
redshift ranges $2 < z < 3$ and $3 < z < 4$ in comparison. The
table lists the number of spectra averaged, the mean redshift, the average 
luminosity at 1500\,\AA{} (see Sect.~\ref{meanspec_all}), the Ly$\alpha$ 
equivalent width, the continuum slope $\beta$, the mean equivalent width of 
low ionisation interstellar absorption lines (see Fig.~\ref{fig_LIS_Lya}) 
and the EWs of C\,IV\,$\lambda\lambda\,1548,1550$, He\,II\,$\lambda\,1640$ 
and C\,III]\,$\lambda\lambda\,1907,\,1909$. All errors given are mean errors 
(see Sect.~\ref{calcmean}).} 
\label{tab_ew}
\centering
\begin{tabular}{c c c}
\hline
\noalign{\smallskip}
$z$ range & $2 - 3$ & $3 - 4$ \\
\noalign{\smallskip}
\hline
\noalign{\smallskip}
$N$ & $42$ & $22$ \\
$z$ & $2.39 \pm 0.03$ & $3.33 \pm 0.04$ \\
$\log L_{1500} [{\rm W/\AA}]$ & $34.14 \pm 0.03$ & $34.35 \pm 0.04$ \\
$W_{\rm Ly\alpha}$ [\AA] & $6.90 \pm 2.10$ & $-10.20 \pm 6.10$ \\
$\beta$ & $-0.56 \pm 0.11$ & $-1.79 \pm 0.13$ \\
$W_{\rm LIS}$ [\AA] & $1.98 \pm 0.10$ & $1.55 \pm 0.15$ \\
$W_{\rm C\,IV}$ [\AA] & $3.80 \pm 0.25$ & $2.33 \pm 0.35$ \\
$W_{\rm C\,III]}$ [\AA] & $-1.10 \pm 0.30$ & $-3.40 \pm 0.70$ \\
$W_{\rm He\,II}$ [\AA] & $-1.30 \pm 0.20$ & $-2.20 \pm 0.40$ \\
\noalign{\smallskip}
\hline
\end{tabular}
\end{table}

To investigate evolutionary effects in our sample we computed in the next 
step mean spectra of the galaxies within the redshift intervals $2 < z < 3$ 
and $3 < z < 4$. The resulting composite spectra with mean redshifts 
$\langle z \rangle = 2.39$ ($\sigma_z = 0.19$) and 
$\langle z \rangle = 3.33$ ($\sigma_z = 0.20$) are based on 42 and 22 
objects, respectively. The results are given in Fig.~\ref{fig_mean_z} and 
Table~\ref{tab_ew}.

Following the normal spectroscopic convention in Table~\ref{tab_ew} (and 
throughout this paper) for net absorption lines we give positive equivalent
width values, while net emission features are characterised by negative 
EW features. The EW mean errors were estimated on the basis of the mean 
errors of the composite spectra, which were derived from the scatter of the 
continuum-subtracted individual spectra, and the uncertainties concerning 
the continuum level determination. In contrast, the mean errors of the 
continuum slope $\beta$ (calculated in the range 1200 to 1800\,\AA{} assuming 
$f(\lambda) \propto \lambda^{\beta}$ following Leitherer et al. 
(\cite{LEI02})) were derived from the scatter of the values measured in the 
individual galaxy spectra (see Fig.~\ref{fig_betaz}). Usually the $\beta$ of 
the mean spectrum and the average $\beta$ of the individual spectra were 
roughly the same, differing by less than $0.1$, which shows that the 
composite spectra are characteristic for the respective samples, at least 
with regard to the continuum slope.

As shown by Fig.~\ref{fig_mean_z} and Table~\ref{tab_ew}, the $2 < z < 3$ 
mean spectrum has a rather flat continuum ($\beta = -0.56 \pm 0.11$) and 
Ly$\alpha$ emission is weak ($W_{\rm Ly\alpha} = +7 \pm 2$\,\AA{}). In 
contrast, the $3 < z < 4$ mean spectrum shows a steep UV continuum 
($\beta = -1.79 \pm 0.13$) and strong Ly$\alpha$ emission 
($W_{\rm Ly\alpha} = -10 \pm 6$\,\AA{}). The equivalent widths of the 
majority of spectral lines and in particular all interstellar absorption 
lines show minor differences only. However, in general, the EWs tend to be 
higher in the lower redshift range. A strong redshift dependence is observed
for the C\,IV resonance doublet, which shows a weaker absorption at higher 
redshifts (as noted already by Mehlert et al. \cite{MEH02}). Furthermore, 
C\,III] displays a stronger emission in the $3 < z < 4$ mean spectrum. 
However, due to the line's location in the region of strong OH bands at 
$z > 3$ the $3 < z < 4$ spectrum contains only five spectra at C\,III]. 
Therefore, the observed increase of C\,III] with redshift may not be 
statistically meaningful.

The composite spectrum of the higher redshift range ($3 < z < 4$) includes 
only few objects with luminosities lower than $\log L_{1500} = 34.2$ (3 out 
of 22; see Fig.~\ref{fig_luvz}), whereas 28 out of 42 objects in the range 
$2 < z < 3$ are below this value. To check the effect of the luminosity 
difference on the spectral properties, we also calculated composite spectra 
with $\log L_{1500} = 34.2$ as lower luminosity limit. The results showed 
that the luminosity does not affect the spectral differences, except perhaps 
for a slight decrease of the Ly$\alpha$ strength in the high-luminosity 
subsample (from $-10 \pm 7$\,\AA{} to $-5 \pm 6$\,\AA{}), caused by the 
omission of two strong Ly$\alpha$ emitters.

\subsection{Comparison with the composite spectrum of Shapley et al.}
\label{comp_lbg}

\begin{figure*}
\centering 
\includegraphics[width=17.8cm,clip=true]{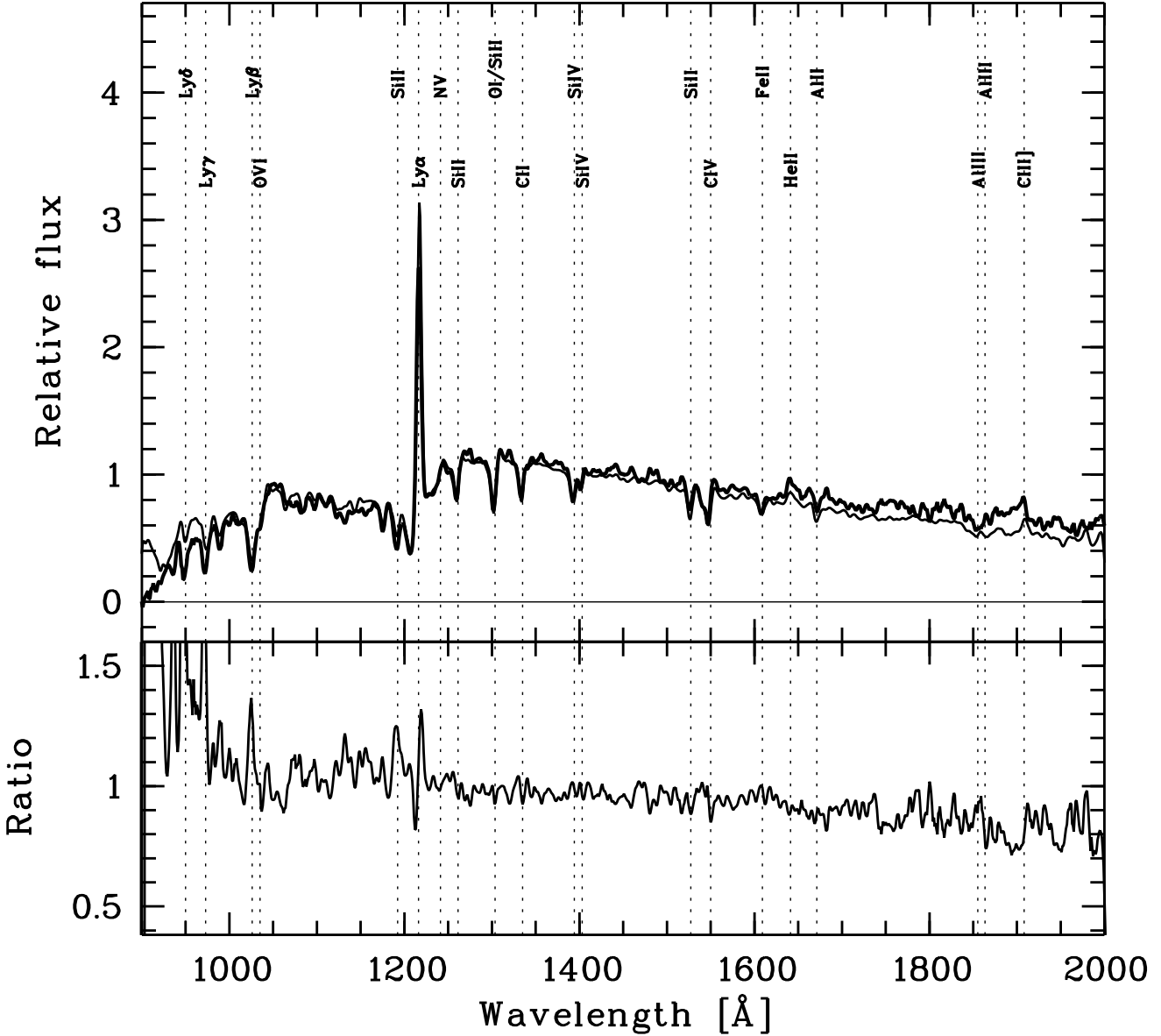}
\caption[]{Comparison of composite spectra representative of the FORS Deep 
Field in the redshift range $2.5 < z < 3.5$ (thick line) and the S03 sample 
at $z \sim 3$ described in Steidel et al. (\cite{STE03}) and \cite{SHA03} 
(thin line). The original S03 mean spectrum was smoothed to the same spectral 
resolution and normalised to the same mean continuum level in the interval 
$1250 - 1500$\,\AA{}. The lower panel gives the ratio between the FDF and the 
S03 composite spectra.}
\label{fig_mean_zLBG}
\end{figure*}

S03 have published a composite spectrum of a homogeneous sample of 
almost 1000 Lyman-break galaxies at $z \sim 3$ selected via the Lyman-break 
technique (see Steidel et al. \cite{STE03}). Because of the different 
selection criteria, the redshift distributions of the FDF galaxies 
($\langle z \rangle = 2.72$, $\sigma_z = 0.49$) and the S03 sample 
($\langle z \rangle = 2.96$, $\sigma_z = 0.29$) differ. 53\,\% of the FDF 
high-redshift galaxies are at $z < 2.5$, whereas this redshift range is 
essentially absent in the S03 sample. Hence, in view of the results of 
Sect.~\ref{meanspec_z}, differences in the spectral properties are to be 
expected. Therefore we calculated a composite spectrum of 28 FDF galaxies 
between $z = 2.5$ and $3.5$ with a similar redshift coverage 
($\langle z \rangle = 3.11$, $\sigma_z = 0.28$) as the S03 sample. 
Fig.~\ref{fig_mean_zLBG} compares this composite spectrum with the S03 
spectrum. The Ly$\alpha$ strength ($W_{\rm Ly\alpha} = -9 \pm 5$\,\AA{}) as 
well as the continuum slope ($\beta = -1.54 \pm 0.13$) of the 
$2.5 < z < 3.5$ spectrum show good agreement with the S03 values 
($W_{\rm Ly\alpha} = -11$\,\AA{} and $\beta = -1.81$). This shows that 
samples with comparable redshift distributions have similar average spectra 
independent of the selection method. 

The fact that the average apparent brightness of the S03 sample 
($R_{\rm Vega} \approx 24.38$\,mag, converted from $R_{\rm AB}$) is somewhat 
lower than that of the $2.5 < z < 3.5$ FDF sample 
($R_{\rm Vega} = 24.19 \pm 0.11$\,mag) apparently has no effect on the 
average spectra. This is confirmed by a comparison of the composite spectra 
of the bright ($\langle R \rangle = 23.80$\,mag) and the faint half 
($\langle R \rangle = 24.58$\,mag) of the $2.5 < z < 3.5$ sample, which 
show very similar spectra. The only possible difference may be a moderately
higher Ly$\alpha$ emission of $-16 \pm 10$\,\AA{} for the faint subsample in 
comparison to $-3 \pm 4$\,\AA{} for the bright subsample.

\subsection{Relations between Ly$\alpha$ emission and other spectral 
properties}
\label{meanspec_lya}

\begin{figure}
\centering 
\includegraphics[width=8.8cm,clip=true]{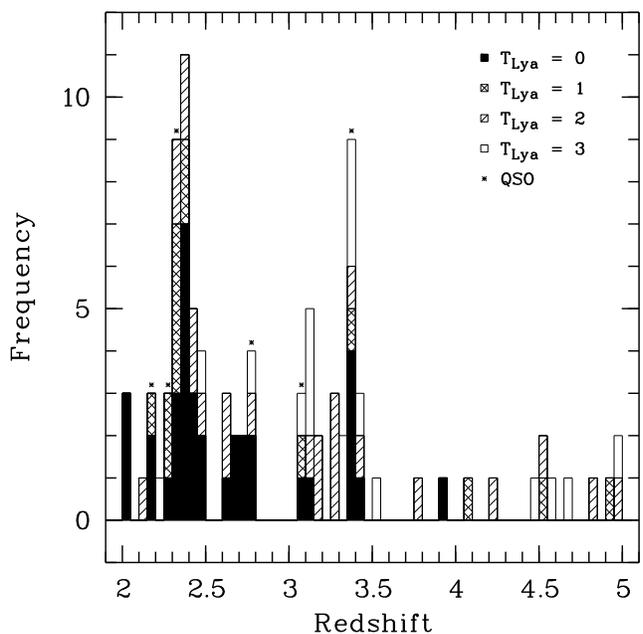}
\caption[]{Redshift distribution of FDF galaxies with different Ly$\alpha$ 
types (see legend). The presence of a quasar at the corresponding redshift 
is marked by an asterisk.}
\label{fig_Tlyaz}
\end{figure}

\begin{table}
\caption[]{Distribution of Ly$\alpha$ types $T_{\rm Ly\alpha}$ 
(as defined in Sect.~\ref{meanspec_lya}) among the FDF high-redshift 
galaxies ($2 < z < 5$). For each Ly$\alpha$ class the total number of 
objects $N$, the median redshift $\overline{z}$, and the mean redshift 
$\langle z \rangle$ with its standard deviation $\sigma_z$ are given.}
\label{tab_TLya}
\centering
\begin{tabular}{c c c c c}
\hline
\noalign{\smallskip}
$T_{\rm Ly\alpha}$ & $N$ & $\overline{z}$ & $\langle{}z\rangle$ & $\sigma_z$ 
\\
\noalign{\smallskip}
\hline
\noalign{\smallskip}
0 &  36 & 2.4 & 2.6 & 0.5 \\ 
1 &  14 & 2.3 & 2.9 & 0.9 \\
2 &  24 & 3.1 & 3.1 & 0.8 \\
3 &  18 & 3.3 & 3.5 & 0.8 \\
\noalign{\smallskip}
\hline
\end{tabular}
\end{table}

\begin{figure*}
\centering 
\includegraphics[width=17.8cm,clip=true]{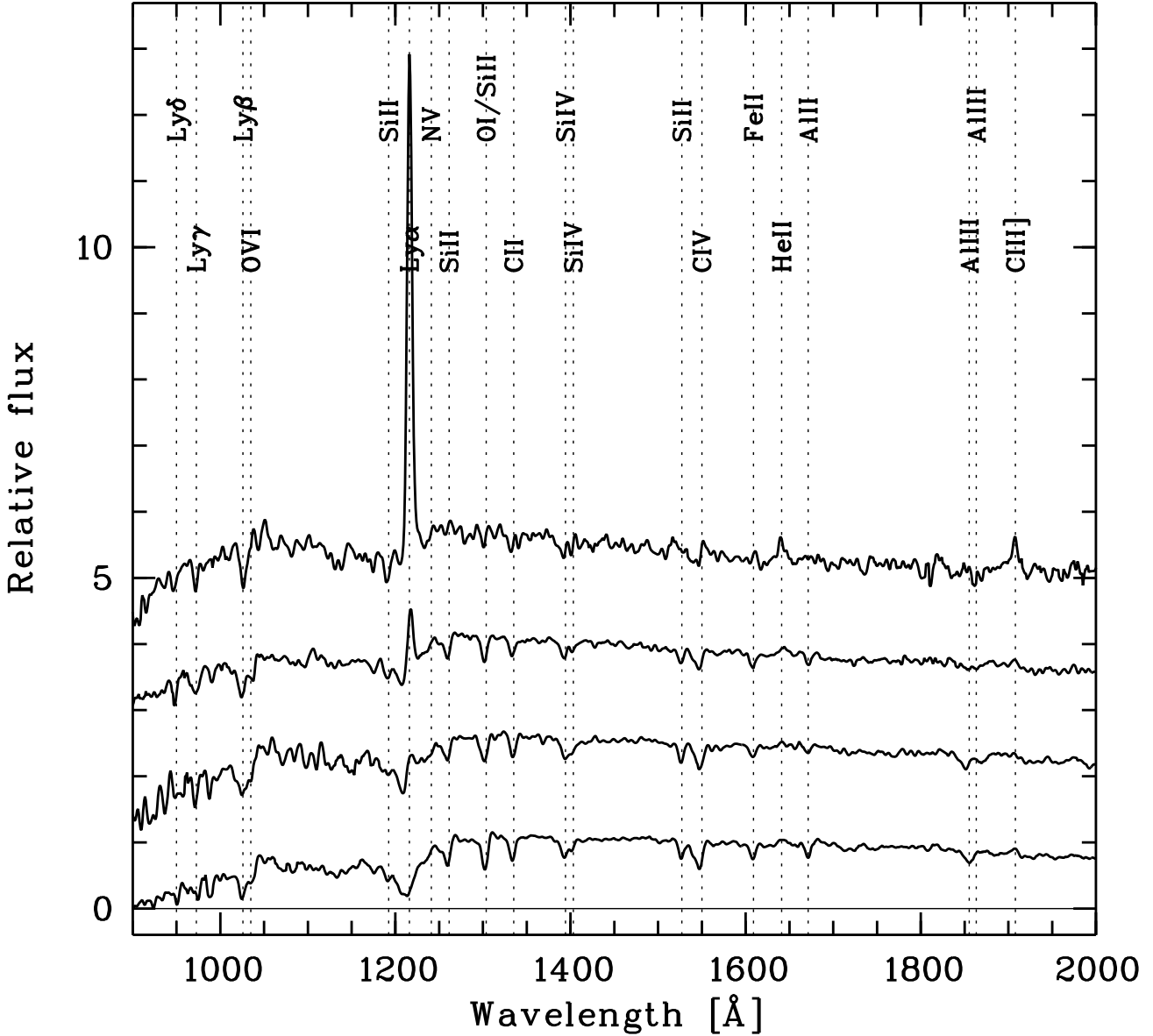}
\caption[]{Comparison of FDF composite spectra representative of different
Ly$\alpha$ types ranging from strong emission ($T_{\rm Ly\alpha} = 3$, 
uppermost spectrum) to pure absorption ($T_{\rm Ly\alpha} = 0$, lowest 
spectrum). The spectra are normalised and shifted by multiples of $1.0$ 
relative flux units.}
\label{fig_mean_lya}
\end{figure*}

\begin{figure}
\centering 
\includegraphics[width=8.8cm,clip=true]{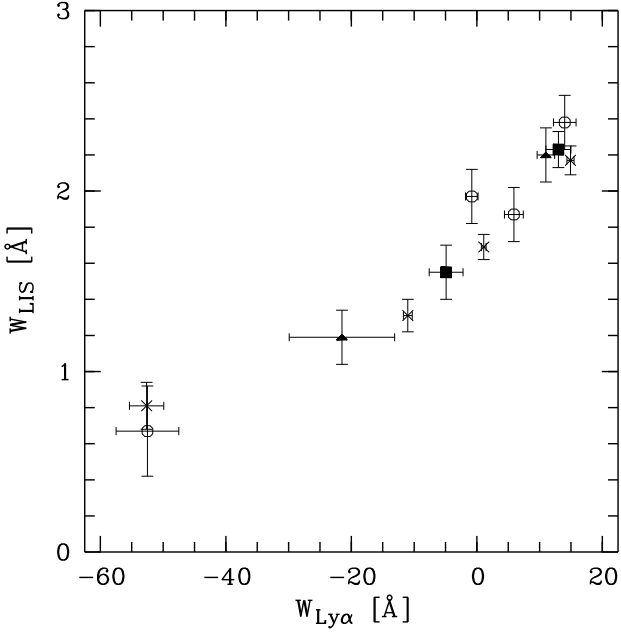}
\caption[]{The dependence of the strength of low-ionisation lines on the 
Ly$\alpha$ equivalent width. $W_{\rm LIS}$ is the average EW of the six 
strong low-ionisation features Si\,II\,$\lambda\,1260$, 
OI/SiII\,$\lambda\,1303$, C\,II\,$\lambda\,1334$, Si\,II\,$\lambda\,1526$, 
Fe\,II\,$\lambda\,1608$, and Al\,II\,$\lambda\,1670$. The individual symbols 
represent the values measured in the FDF composite spectra for different 
Ly$\alpha$ strengths {\em and} redshift ranges. Open circles show the values 
for the redshift range $2 < z < 4$. Filled squares and triangles represent 
the smaller ranges $2 < z < 3$ and $3 < z < 4$, respectively. The 
$1\,\sigma$ error bars include the variance between the individual spectra 
and the errors of the continuum level definition. For comparison the 
corresponding data from \cite{SHA03} are also included (crosses).}
\label{fig_LIS_Lya}
\end{figure}

\begin{figure}
\centering 
\includegraphics[width=8.8cm,clip=true]{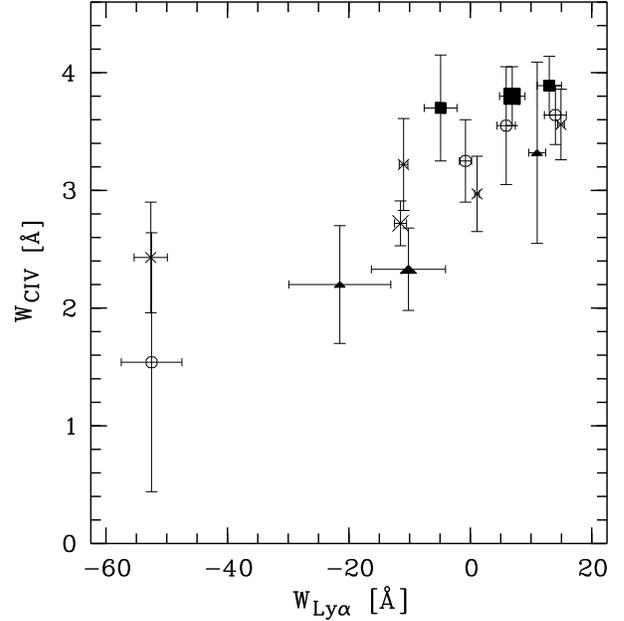}
\caption[]{The dependence of the strength of C\,IV on the Ly$\alpha$ 
equivalent width. The smaller symbols are as in Fig.~\ref{fig_LIS_Lya}. The
big filled square, filled triangle, and cross show the EWs measured in the 
$2 < z < 3$, $3 < z < 4$, and the S03 composite spectrum, respectively.}
\label{fig_CIV_Lya}
\end{figure}

\begin{figure}
\centering 
\includegraphics[width=8.8cm,clip=true]{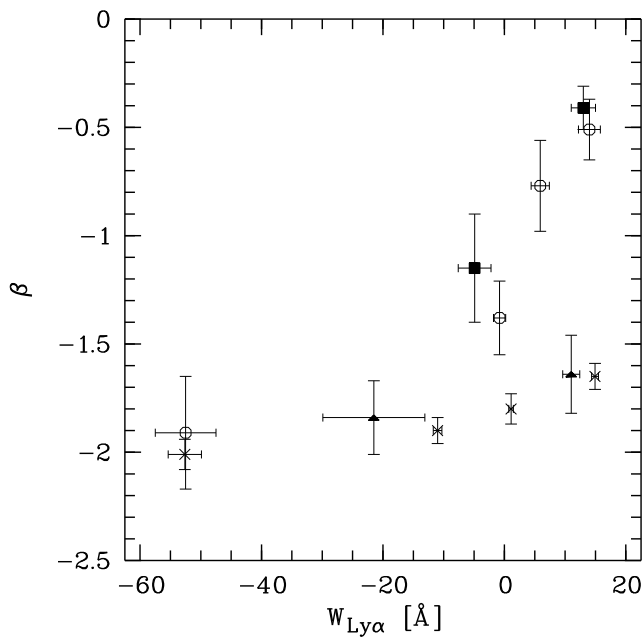}
\caption[]{The dependence of the strength of the continuum slope parameter
$\beta$ (see Leitherer et al. \cite{LEI02}) on the Ly$\alpha$ equivalent 
width. Symbols are as in Fig.~\ref{fig_LIS_Lya}. The $\beta$ uncertainties 
were obtained from the scatter of the values derived for the individual 
spectra. The values given in \cite{SHA03} were calculated from the colour 
excess $E_{\rm B-V}$, whereas the FDF values were derived by fitting the 
composite spectra. Hence, the S03 values (crosses) were corrected assuming a 
constant offset. The shift ($\Delta\beta = -0.92 \pm 0.05$) was estimated 
from the $\beta$ measured in the published composite spectrum of the total 
S03 sample.}  
\label{fig_beta_Lya}
\end{figure}

\begin{figure}
\centering 
\includegraphics[width=8.8cm,clip=true]{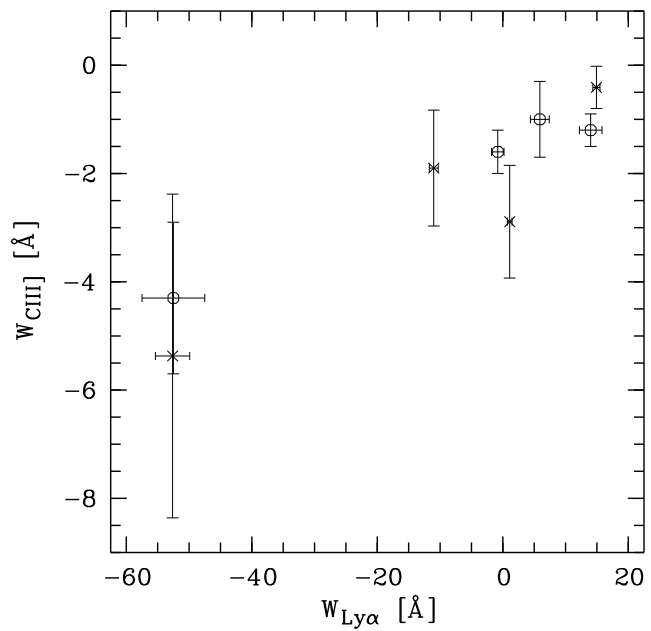}
\caption[]{The dependence of the strength of 
C\,III]\,$\lambda\lambda\,1907,\,1909$ on the Ly$\alpha$ equivalent width. 
The open circles represent the values measured in the FDF composite spectra 
for different Ly$\alpha$ strengths. For comparison the corresponding data 
from \cite{SHA03} are also included (crosses).}
\label{fig_CIII_Lya}
\end{figure}

\begin{figure}
\centering 
\includegraphics[width=8.8cm,clip=true]{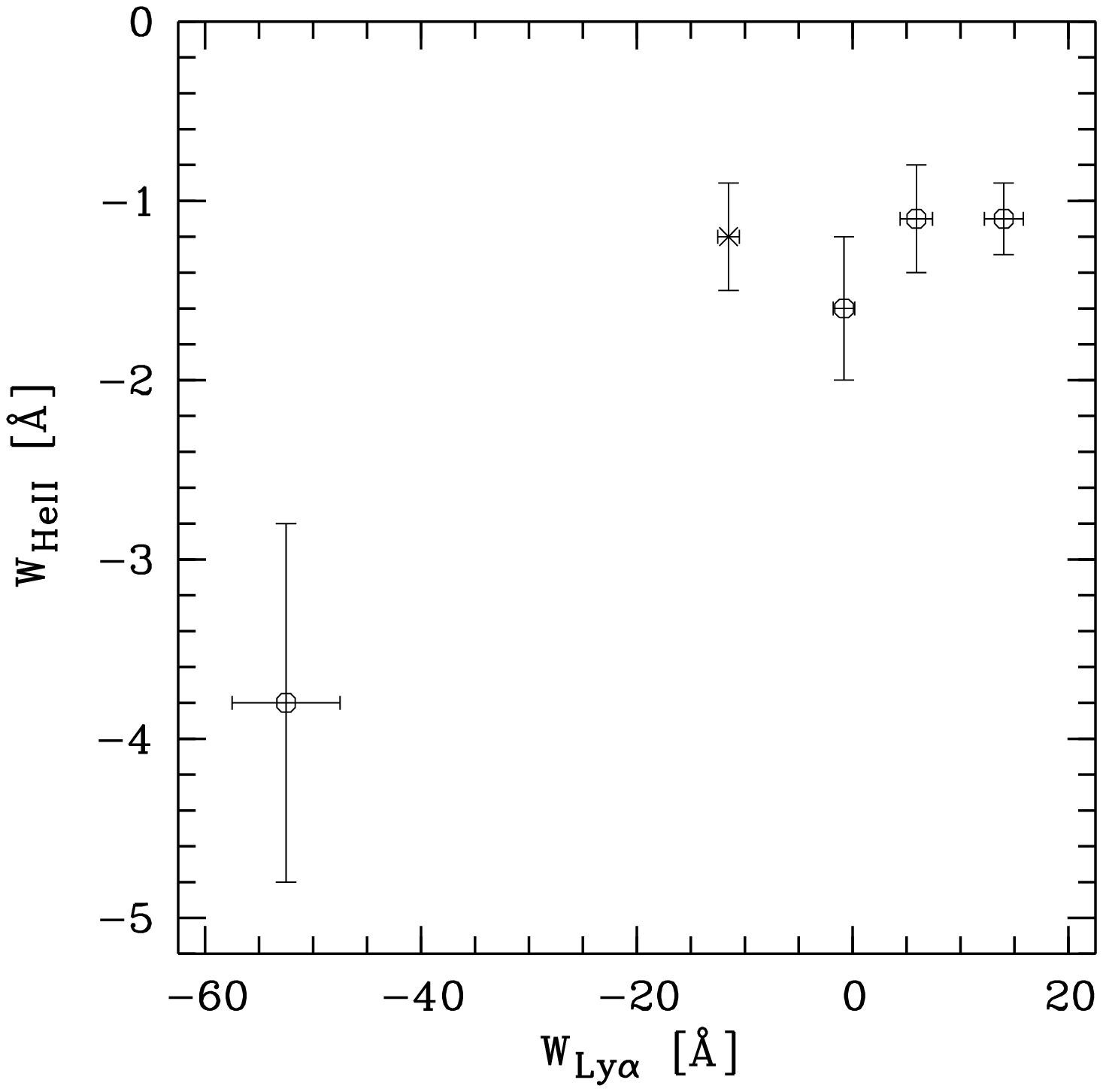}
\caption[]{The dependence of the strength of He\,II\,$\lambda\,1640$ on the 
Ly$\alpha$ equivalent width. The open circles represent the values measured 
in the FDF composite spectra for different Ly$\alpha$ strengths. The cross 
indicates the He\,II equivalent width of the composite spectrum of the S03 
sample.}
\label{fig_HeII_Lya}
\end{figure}

In part of our spectra the Ly$\alpha$ line occurs in emission. The emission 
originates from recombination of hydrogen ionised either by hot stars or by 
an AGN. The escape of Ly$\alpha$ photons is hampered by multiple resonance 
scattering resulting in large path lengths through the galaxy (see Charlot 
\& Fall \cite{CHA93}). Hence, the Ly$\alpha$ emission is very sensitive to 
the dust content and the geometry and structure of the H\,I velocity field 
(see e.g. Kunth et al. \cite{KUN98}).

To study the relation between the Ly$\alpha$ emission and other spectral 
features we divided the high-redshift galaxy sample spectra into four 
classes according to the EW of the Ly$\alpha$ emission component 
$W_{{\rm Ly\alpha},{\rm e}}$. The four classes ($T_{\rm Ly\alpha}$) were 
defined in the following way:\\
$T_{\rm Ly\alpha} = 0$: No (detectable) emission. \\           
$T_{\rm Ly\alpha} = 1$: Very weak emission (not measurable). \\
$T_{\rm Ly\alpha} = 2$: 
$2\,{\rm \AA} \le -W_{{\rm Ly\alpha},{\rm e}} < 20$\,\AA{}. \\
$T_{\rm Ly\alpha} = 3$: $-W_{{\rm Ly\alpha},{\rm e}} \ge 20$\,\AA{}.

Mean and median redshifts of the individual Ly$\alpha$ classes are listed in 
Table~\ref{tab_TLya}. The resulting redshift distribution of the different 
types is shown in Fig.~\ref{fig_Tlyaz}. The histogram indicates a conspicuous 
evolution of the typical Ly$\alpha$ characteristics from $z = 2$ to 5. At 
redshifts $2 < z < 3$ only 26\,\% of the objects shows significant Ly$\alpha$ 
emission ($T_{\rm Ly\alpha} \ge 2$). Only 3 of 18 Ly$\alpha$-bright galaxies 
(LABs, $T_{\rm Ly\alpha} = 3$) are at $z < 3$. They represent 6\,\% of all 
galaxies in the range $2 < z < 3$. At higher redshifts the fraction of 
galaxies with $T_{\rm Ly\alpha} \ge 2$ increases to 67\,\% at redshifts 
$3 < z < 4$ and 73\,\% at $4 < z < 5$. Considering the galaxies with strong 
Ly$\alpha$ emission only, the corresponding values are 37\,\% and 36\,\%. 
Hence, the evolution of the average Ly$\alpha$ strength between $z = 3.3$ and 
$2.4$ is caused by a change of the frequency {\em and} the strength of 
Ly$\alpha$ emission, if present.  

For the analysis of the correlation of Ly$\alpha$ strength with other 
spectral features, we again took the 64 galaxy spectra in the range 
$2 < z < 4$ and calculated composite spectra for each of the four Ly$\alpha$ 
classes introduced above. In Fig.~\ref{fig_mean_lya} the resulting spectra 
are plotted. To investigate the redshift dependence of the correlations, four 
further composite spectra were calculated, differing in redshift range 
($2 < z < 3$ and $3 < z < 4$) and Ly$\alpha$ class $T_{\rm Ly\alpha}$ 
($0 - 1$ and $2 - 3$). To avoid too small numbers of spectra in individual 
composite spectra, the Ly$\alpha$ types were grouped into two categories 
only. For all composite spectra EWs and continuum slopes were measured, 
allowing us to check whether the FDF sample follows the relations found by 
\cite{SHA03}. 

As pointed out by Pettini et al. (\cite{PET00}) and \cite{SHA03}, prominent 
low-ionisation absorption lines of interstellar origin in spectra of 
high-redshift galaxies (like OI/SiII\,$\lambda\,1303$ and 
Si\,II\,$\lambda\,1526$) are normally saturated. The EWs of such lines are 
mainly determined by the product of the covering fraction of the neutral 
gas clouds and the velocity field, which makes them suitable indicators of 
the geometry and kinematics of neutral gas in high-redshift galaxies.
Fig.~\ref{fig_LIS_Lya} shows the average EW of six prominent low-ionisation 
features as a function of the Ly$\alpha$ equivalent width $W_{\rm Ly\alpha}$. 
The four data points of the different Ly$\alpha$ types (open circles) 
indicate a significant decrease of $W_{\rm LIS}$ with increasing 
$W_{\rm Ly\alpha}$. The spectrum representing the strong Ly$\alpha$ emitters 
shows a $W_{\rm LIS}$ more than three times weaker than does the spectrum 
with pure Ly$\alpha$ absorption, in close agreement with the results of 
Shapley et al. (crosses). Moreover, significant differences of the relations 
for $2 < z < 3$ (filled squares) and $3 < z < 4$ (filled triangles) cannot 
be detected, which suggests that a redshift dependence, if present, must be 
weak. Consequently, Fig.~\ref{fig_LIS_Lya} suggests a strong dependence of 
the Ly$\alpha$ strength on the geometry and kinematics of the H\,I clouds 
(in accordance with Kunth et al. \cite{KUN98}) independent of the redshift.

Fig.~\ref{fig_CIV_Lya} shows C\,IV\,$\lambda\lambda\,1548,\,1550$ as a
function of the Ly$\alpha$ strength. As C\,IV originates primarily in 
photospheres and winds of luminous hot stars (Walborn et al. \cite{WAL95}; 
Heckman et al. \cite{HEC98}) and the wind power depends on the metallicity, 
$W_{\rm C\,IV}$ is a good metallicity indicator (Leitherer et al. 
\cite{LEI01}). The figure confirms the strong dependence of the C\,IV 
strength on the redshift, noted already by Mehlert et al. (\cite{MEH02}). On 
the other hand, only a weak dependence of $W_{\rm C\,IV}$ on 
$W_{\rm Ly\alpha}$ is indicated. The large $W_{\rm C\,IV}$ mean errors  
suggest a marked scatter of the C\,IV strength for each Ly$\alpha$ type. 
Consequently, the Ly$\alpha$ emission appears to be rather independent of 
metallicity-dependent stellar wind properties.

In Fig.~\ref{fig_beta_Lya} we investigate the dependence of the continuum 
slope $\beta$ on $W_{\rm Ly\alpha}$. As discussed in 
Sect.~\ref{meanspec_beta}, $\beta$ essentially traces the attenuation of
hot star continua by interstellar dust. The diagram reveals a conspicuous
redshift dependence of the $\beta - W_{\rm Ly\alpha}$ relation. The galaxies 
at $z < 3$ (filled squares), which represent the main contribution to the FDF 
$2 < z < 4$ sample (open circles), show a strong dependence of the continuum 
slope and Ly$\alpha$. In contrast, the galaxies at $z > 3$ (filled triangles) 
indicate a rather weak $\beta - W_{\rm Ly\alpha}$ correlation, resembling the 
relation for the S03 sample (crosses). The S03 $\beta$ values had to be 
corrected (see Fig.~\ref{fig_beta_Lya}) to allow a comparison with the FDF 
values. Therefore, we computed $E_{\rm B-V}$ following \cite{SHA03} (using 
Starburst99 of Leitherer et al. (1999) and assuming the Calzetti law for dust 
extinction, solar metallicity, and an underlying stellar population with 
300\,Myr of constant star-formation) to preclude possible systematic 
adjustment errors. The resulting $E_{\rm B-V}$ values confirm the trend 
indicated by Fig.~\ref{fig_beta_Lya}. 

According to \cite{SHA03} the $\beta - W_{\rm Ly\alpha}$ relation can be 
explained assuming that starburst regions are covered by dusty gas clouds, 
which cause a decrease of the escape probability of Ly$\alpha$ photons (see 
above) {\em and} an increase of the extinction of the stellar continua by 
dust. To allow significant Ly$\alpha$ emission from galaxies at $z < 3$, the 
increase of the dust content in luminous galaxies at lower redshifts must not 
lead to a strong decrease of the escape probability of Ly$\alpha$ photons. 
Hence, the observed evolution of the $\beta - W_{\rm Ly\alpha}$ relation 
requires that the amount of neutral gas grows more slowly than the dust 
content with decreasing redshift and/or the typical geometric distribution 
and the velocity field of the interstellar medium have to change suitably.   

The nebular emission feature C\,III]\,$\lambda\lambda\,1907,\,1909$ traces
the electron temperature in H\,II regions, which depends on the composition
of the stellar population and the metallicity (Heckman et al. \cite{HEC98}).
Fig.~\ref{fig_CIII_Lya} shows that the FDF galaxies (open circles) follow 
well the relation between Ly$\alpha$ and C\,III] observed in the S03 sample, 
indicating strong C\,III] emission, and thus hot H\,II regions in the case 
of Ly$\alpha$-bright galaxies. 

Finally, we investigated the dependence of the equivalent width of the 
He\,II\,$\lambda\,1640$ emission feature on $W_{\rm Ly\alpha}$. 
He\,II\,$\lambda\,1640$ can form only if a hard continuum is present. 
Excluding AGNs, significant He\,II emission is expected in the presence of 
Wolf-Rayet stars only (Schaerer \cite{SCHA03}). Since Wolf-Rayet stars 
require a young starburst (age interval $3 < t < 7$\,Myr for solar 
metallicity), He\,II is an important indicator of the recent star-formation 
history and the metallicity. Fig.~\ref{fig_HeII_Lya} shows that for the 
majority of FDF spectra the EW of He\,II is close to $-1$\,\AA{}. This result 
agrees well with $W_{\rm He\,II}$ derived for the S03 composite spectrum for 
the whole sample (cross). On the other hand, an inspection of the individual 
FDF spectra show particularly large variations of the He\,II feature. As an 
example we note that FDF-5903 (Fig.~\ref{fig_galspec}(c)) has a significantly 
stronger He\,II emission ($W_{\rm He\,II} = -1.6 \pm 0.2$\,\AA{}) than the 
average while other galaxies show He\,II emission statistically significantly 
below the average. Furthermore, Fig.~\ref{fig_HeII_Lya} suggests that strong 
Ly$\alpha$ emitters have particularly high $W_{\rm He\,II}$. However, because 
of the large error in $W_{\rm He\,II}$ of the $W_{\rm Ly\alpha} = -53$\,\AA{} 
point, this result is not statistically significant. These observations may 
indicate that, like in the local universe, there are high-redshift 
`Wolf-Rayet galaxies' having exceptionally young starburst populations.

\subsection{Relations with the continuum slope $\beta$}\label{meanspec_beta}  

\begin{figure}
\centering 
\includegraphics[width=8.8cm,clip=true]{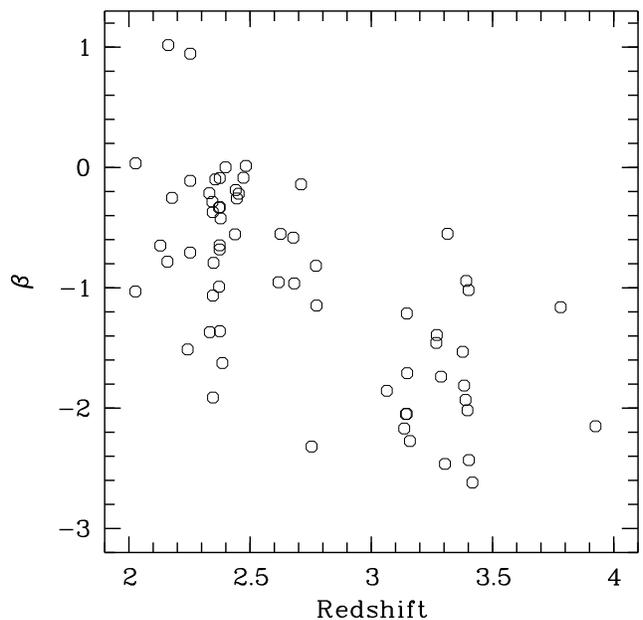}
\caption[]{UV slope $\beta$ between 1200 and 1800\,\AA{} (Leitherer et al. 
\cite{LEI02}) as a function of redshift for the FDF galaxy subsample used in 
the calculation of the composite spectra (see Sect.~\ref{meanspec_all}). 
Typical errors of $\beta$ are around $0.1$.}
\label{fig_betaz}
\end{figure}

\begin{figure}
\centering 
\includegraphics[width=8.8cm,clip=true]{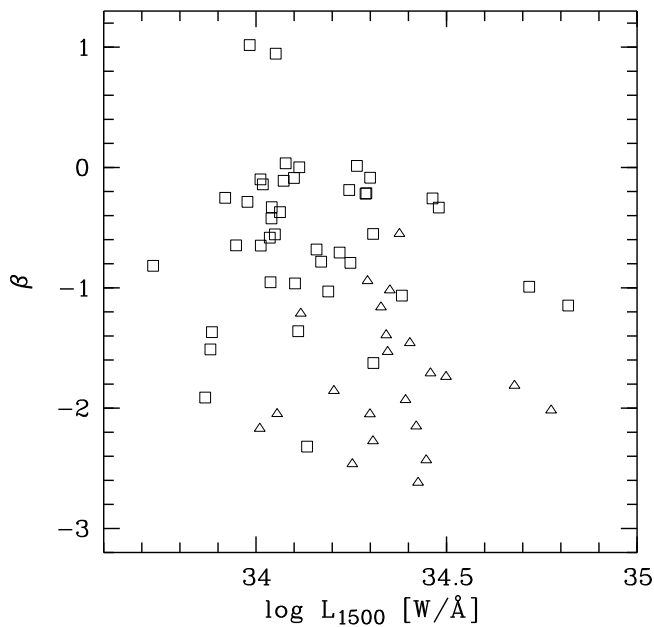}
\caption[]{UV slope $\beta$ as a function of the luminosities at 1500\,\AA{} 
for the FDF galaxy subsample used in the calculation of the composite 
spectra. Squares mark galaxies at $z < 3$. Triangles indicate galaxies at 
$z > 3$. Typical errors of $\beta$ are around $0.1$.}
\label{fig_beta_luv}
\end{figure}

\begin{figure}
\centering 
\includegraphics[width=8.8cm,clip=true]{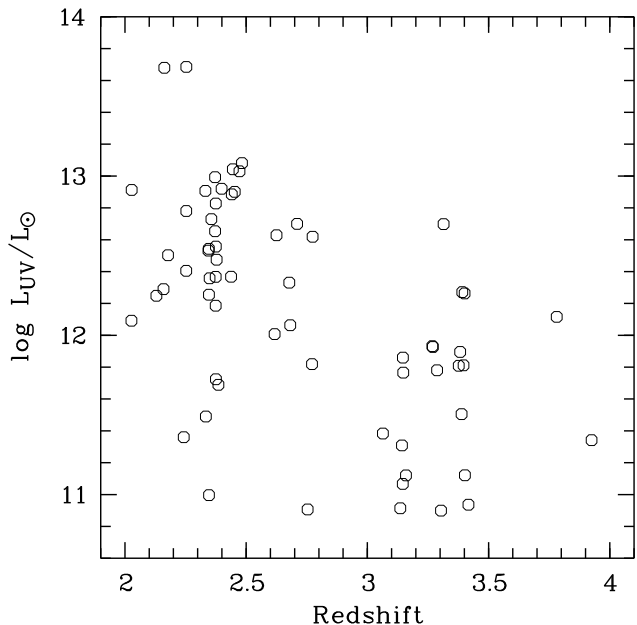}
\caption[]{Total extinction-corrected UV luminosity in solar units as a 
function of redshift for the FDF galaxy subsample used in the calculation of 
the composite spectra.}
\label{fig_cluvz}
\end{figure}

\begin{figure}
\centering 
\includegraphics[width=8.8cm,clip=true]{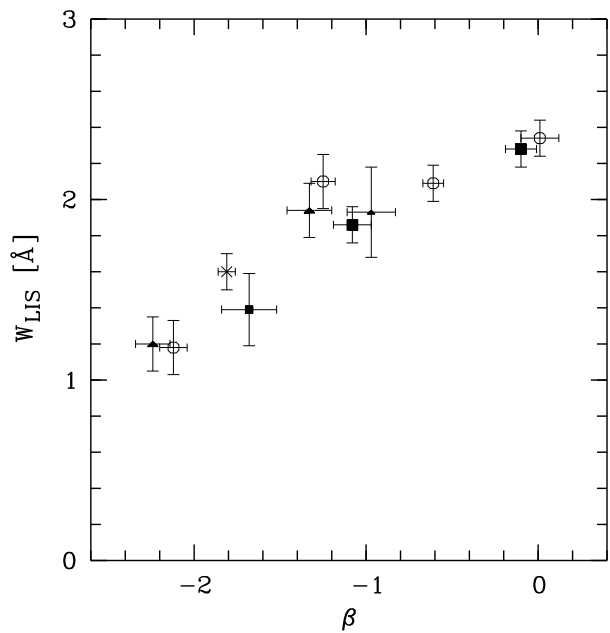}
\caption[]{The dependence of the strength of low ionisation lines (see 
Fig.~\ref{fig_LIS_Lya}) on the continuum slope $\beta$. The open circles 
indicate values measured in FDF composite spectra representative of the 
different quartiles (containing 16 objects each) of the $\beta$ distribution 
derived from the individual spectra. The subsamples of the ranges 
$2 < z < 3$ (filled squares) and $3 < z < 4$ (filled triangles) were divided 
into two parts of equal number and averaged using $\beta$ as classification 
parameter. Data points representing the six steepest continua 
($\beta < -1.3$) at $z < 3$ and the five flattest continua ($\beta > -1.3$) 
at $z > 3$ are marked by a small filled square and triangle, respectively. 
For comparison the corresponding data from \cite{SHA03} are also shown 
(crosses).}  
\label{fig_LIS_beta}
\end{figure}

\begin{figure}
\centering 
\includegraphics[width=8.8cm,clip=true]{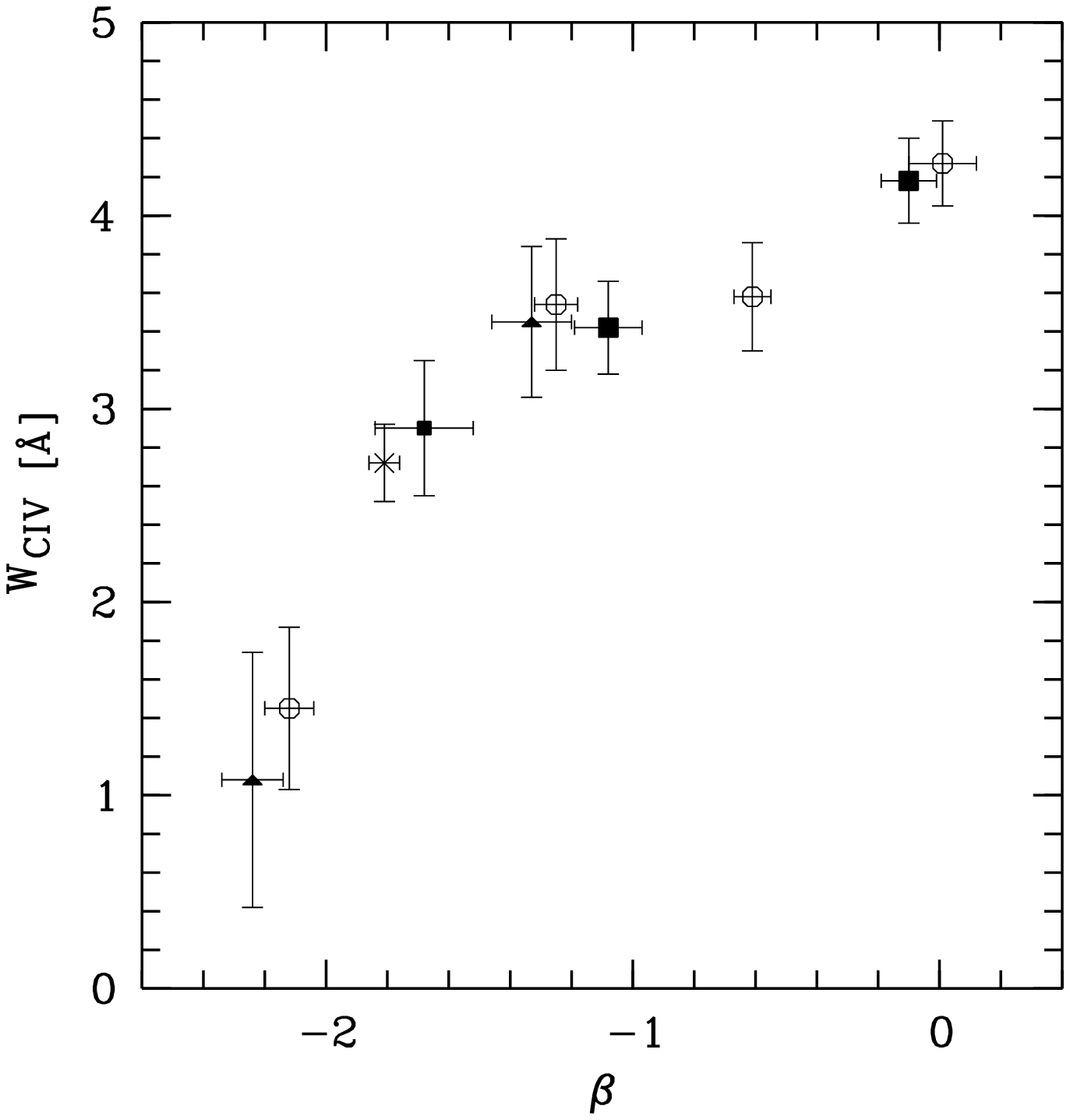}
\caption[]{The dependence of the C\,IV equivalent width on the continuum 
slope parameter $\beta$. The data points were derived from composite spectra 
for different $\beta$ and $z$ intervals. Symbols are as in 
Fig.~\ref{fig_LIS_beta}.} 
\label{fig_CIV_beta}
\end{figure}

In principle, the UV continuum slope $\beta$ can be affected by various 
effects, such as the star-formation history (which determines the composition 
of the stellar population), the initial mass function, the chemical 
composition, dust extinction and the contribution of a possible AGN. However, 
for starburst galaxies showing a very young stellar population the 
predominant effect is expected to be dust reddening (see Calzetti et al. 
\cite{CAL94}; Leitherer et al. \cite{LEI99}, \cite{LEI02}). This implies that 
$\beta$ depends on the distribution of dust in the galaxy and on the 
metallicity (see Heckman et al. \cite{HEC98}), which has an important impact 
on the dust formation.  

In Table~\ref{tab_ew} $\beta$ was found to be strongly redshift dependent, 
evolving from $\beta = -1.8$ at $z \sim 3.3$ to $\beta = -0.6$ at 
$z \sim 2.4$. To study this effect in more detail we also calculated the 
$\beta$ values of the individual galaxies. The resulting 
Fig.~\ref{fig_betaz} confirms the change of $\beta$ with redshift. 

To exclude that these results are due to a luminosity-related selection 
effect, we plotted in Fig.~\ref{fig_beta_luv} $\beta$ as a function of the 
galaxy luminosity at 1500\,\AA{} (see Fig.~\ref{fig_luvz}). As shown by the 
figure, the $z > 3$ objects show a distinctly higher average luminosity 
($\langle \log L_{1500} [{\rm W/\AA}] \rangle = 34.35 \pm 0.04$ in comparison 
to $34.14 \pm 0.03$ for the lower redshift range). This effect is mainly due 
to our selection criteria, causing an increase of the luminosity limit with 
increasing redshift. Thus, we cannot exclude the existence of reddened faint 
$z > 3$ galaxies. On the other hand, the $z < 3$ sample contains only a few 
`blue' galaxies, although the brightness limits would allow their detection. 
Furthermore, a different sample selection would not increase the number of 
such galaxies significantly, since only a few luminous FDF objects have 
suitable magnitudes and colours. Finally, the $I$ magnitude based sample 
selection does not introduce a significant $\beta$-related selection effect. 
Assuming no evolution, the ratio of `blue' to 'red' galaxies should be almost 
the same for $z \sim 2.3$ and $z \sim 3.3$, since 
$I(z \sim 3.3) - I(z \sim 2.3)$ 
$\approx f(1800\,{\rm \AA}) / f(2400\,{\rm \AA})$ is similar for both galaxy 
types (compare Type IV and V in Fig.~\ref{fig_sed}). Consequently, the 
conspicuous difference between the mean $\beta$ of the 
$\log L_{1500} [{\rm W/\AA}] \ge 34.2$ objects 
($\langle\beta\rangle = -0.58 \pm 0.13$ for $z < 3$ in comparison to 
$\langle\beta\rangle = -1.74 \pm 0.13$ for $z > 3$) is obviously real and 
cannot be explained by the selection procedure applied.
   
The very different continuum slopes for similar absolute brightnesses suggest
that there is a strong evolution of the intrinsic UV luminosities of the 
galaxies. Hence, we estimated this quantity following Leitherer et al. 
(\cite{LEI02}). Taking 
\begin{equation}\label{eq_A1500}
A_{1500} = 2.19\,(\beta - \beta_0) 
\end{equation}
($A_{1500}$ being the attenuation at 1500\,\AA{} in mag), $\beta_0 = -2.5$ 
and 
\begin{equation}\label{eq_LUV}
\log L_{\rm UV} = \log L_{1500} + 0.4\,A_{1500} + 3.2, 
\end{equation}
we obtained the values plotted in Fig.~\ref{fig_cluvz}. As expected, there is 
a significant redshift dependence of the extinction-corrected total UV 
luminosity, leading to very luminous galaxies 
($L_{\rm UV} > 10^{12}\,{\rm L_{\sun}}$) at $z < 3$, while such objects are 
rare at higher redshifts. The galaxies at $z < 2.5$ with 
$L_{\rm UV} \sim 10^{13}\,{\rm L_{\sun}}$ are particularly interesting, since 
they are as bright as the most luminous star-forming galaxies known, 
implying star-formation rates of several $100\,{\rm M_{\sun}}/{\rm yr}$ and 
more. Fig.~\ref{fig_cluvz} can be explained in the sense that the masses of
bright starbursts grow with cosmic age.     
 
Next, we analysed the dependence of $\beta$ on other spectral properties. For 
this, we divided our sample of 64 galaxies into four equally-occupied $\beta$ 
intervals and constructed composite spectra. Further $\beta$-dependent mean 
spectra were calculated by halving the subsamples of the ranges $2 < z < 3$ 
and $3 < z < 4$ according to the continuum slope. The strong $\beta$ 
evolution made it necessary to construct composite spectra including the six 
steepest spectra ($\beta < -1.3$) with $z < 3$ and the five flattest ones 
($\beta > -1.3$) with $z > 3$, respectively.
 
In Fig.~\ref{fig_LIS_beta} we show the average EW of the six strong 
low-ionisation lines (see Fig.~\ref{fig_LIS_Lya}) as a function of $\beta$. 
The figure indicates a monotonic relation, characterised by a decrease of 
$W_{\rm LIS}$ with bluer continua. A slight redshift dependence of the 
relation cannot be excluded, since $W_{\rm LIS}$ seems to be smaller at 
constant $\beta$ towards lower redshifts. Heckman et al. (\cite{HEC98}) 
conjecture that the $W_{\rm LIS} - \beta$ relation could be caused by a 
relation between dust extinction and turbulent velocity of the interstellar 
medium. This could be due to the dustier galaxies being more massive and 
showing stronger and more violent star-formation. According to \cite{SHA03} 
the $W_{\rm LIS} - \beta$ relation may be due to the change of the covering 
fraction of the hot stars by interstellar clouds consisting of neutral gas 
and dust, which gives a more direct link between both quantities than the 
former explanation. 

Fig.~\ref{fig_CIV_beta} indicates a relatively tight relation between 
$W_{\rm C\,IV}$ and $\beta$, which does not show a redshift evolution. In 
part the strong decrease of $W_{\rm C\,IV}$ with $\beta$ for $\beta < -1.5$ 
can be due to a C\,IV emission component. Fig.~\ref{fig_CIV_beta} supports
the assumption that the dust reddening depends partly on the metallicity, 
which is traced by C\,IV (see Heckman et al. \cite{HEC98}; Leitherer et al. 
\cite{LEI01}; Mehlert et al. \cite{MEH02}).

\section{Implications and conclusions}\label{implications}

As described in the preceding sections, in addition to providing valuable 
information on the spectral properties of distant galaxies, the FDF 
spectroscopic survey also provided interesting new data on the evolution of 
the properties of bright starburst galaxies $2 < z < 5$. The main results on 
this redshift evolution can be summarised as follows:

\begin{itemize}
\item Although the spectra of the starburst galaxies are basically very
similar at all observed redshifts, there is a tendency for the net line 
absorption to become weaker with increasing redshift. In particular, our 
galaxy spectra with $z > 3$ show on average weaker absorption and/or 
stronger emission components. The effect is particularly conspicuous for the 
Ly$\alpha$ line and the C\,IV resonance doublet.
\item The UV continuum slope tends to become flatter with decreasing 
redshift. Since for bright starburst galaxies this slope is mainly 
determined by the internal reddening, this correlation also indicates an 
increase of dust reddening with cosmic age.
\item Our data show an anticorrelation of low-ionisation interstellar 
absorption lines with the Ly$\alpha$ emission strength (reported already by 
\cite{SHA03}) and a positive correlation of these lines with the reddening. 
These correlations do not show a significant redshift dependence.
\item The C\,IV absorption strength (while clearly dependent on the redshift)
shows only little correlation with the Ly$\alpha$ emission or absorption
strength. On the other hand, a positive, redshift-independent correlation 
between the C\,IV absorption strength and the dust reddening was found.
\item The anticorrelation between the reddening and the Ly$\alpha$ emission
strength shows a striking redshift dependence, at least for
absorption-dominated Ly$\alpha$ profiles, where the average UV continuum 
flattens with decreasing redshift.  
\item Although our flux-limited spectroscopic sample is observationally 
biased towards absolutely brighter objects at higher redshifts, the average 
intrinsic UV luminosity of the starburst galaxies in our sample decreases
with redshift for $2 < z < 4$.  
\end{itemize}

As shown by Mehlert et al. (\cite{MEH02}) the increase of the C\,IV 
absorption strength with decreasing redshift can be readily explained by the 
cosmic heavy element enrichment history of starburst galaxies in an evolving 
universe. The cosmic chemical evolution also provides a plausible explanation 
for the observed dependence of the UV continuum slope $\beta$ and the 
Ly$\alpha$ emission strength on the redshift. For a lower metallicity we 
expect a lower dust content (and consequently a lower reddening), a higher 
escape probability for Ly$\alpha$ photons, and higher temperatures of the 
radiation field and gas of the starburst galaxies. This prediction is in good 
agreement with the observed steeper continuum slope and (on average) larger 
Ly$\alpha$ emission observed for $z > 3$. 

On the other hand, the Ly$\alpha$ emission shows a relatively strong 
correlation with the low-ionisation absorption lines as compared to the
relatively poor correlation with the C\,IV absorption. Since $W_{\rm LIS}$
traces the properties of (dusty) interstellar H\,I clouds (see \cite{SHA03}), 
a strong influence of the velocity field and/or the geometric structure of 
the interstellar medium (and resulting covering fractions) on the observed 
Ly$\alpha$ flux and its evolution can be assumed. The observational result 
that a significant fraction of the galaxies with $z > 3$ show Ly$\alpha$ in 
absorption, although the average reddening is distinctly lower than for the 
corresponding $2 < z < 3$ galaxies, could also be explained by an evolution 
of the properties of the interstellar medium. 
  
The decrease of the average Ly$\alpha$ emission and the conspicuous 
enhancement of the dust reddening with decreasing redshift may be partly due 
to an increase of the average cold gas and dust mass of the galaxies, which 
impedes the escape of UV photons from the galaxy by an enhanced obscuration 
of the luminous hot stars and H\,II regions in the starburst cores. 
Assuming that the intrinsic UV luminosities of the high-redshift galaxies 
are characteristic for their starburst mass, the observed increase of the 
mean absolute luminosity with decreasing redshift of the galaxies in our 
$2 < z < 4$ sample suggests mass evolution as well. This result can be 
interpreted within current hierarchical models for the formation and 
evolution of galaxies. 

Summarising, we conclude that the observed evolution with redshift of the 
basic properties of the galaxies in the FDF spectroscopic survey sample can 
be well explained on the basis of present theoretical concepts and theories 
of the formation and chemical evolution of galaxies. Hence, the data 
described in this paper provide strong further support for these theoretical 
concepts and models as well as a potential basis for additional 
investigations of the details of such models.

\begin{acknowledgements}
We thank the Paranal staff for their support. This research was supported by
the German Science Foundation (DFG) (Sonderforschungsbereiche 375 and 439).
\end{acknowledgements}

\end{document}